\documentclass[useAMS,usenatbib]{mn2e}
\pdfoutput=1
\def\etal{{\em et al.}\ }
\usepackage{graphicx}

\title[Model independent means of categorizing X-ray binaries.]{Model independent means of categorizing X-ray binaries.  I: Colour-Colour-Intensity Diagrams}
\author[Saeqa Dil Vrtilek and Bram Seth Boroson]{Saeqa Dil Vrtilek$^{1}$\thanks{E-mail:
svrtilek@cfa.harvard.edu} and Bram Seth Boroson 
$^{2}$
\\
$^{1}$Harvard-Smithsonian Center for Astrophysics, 60 Garden Street, Cambridge, MA 02138, USA\\
$^{2}$Clayton State University, Clayton, Georgia, USA}
\begin{document}

\date{}

\pagerange{\pageref{firstpage}--\pageref{lastpage}} \pubyear{2012}

\maketitle

\label{firstpage}

\begin{abstract}
The diverse behaviors displayed by X-ray binaries make it difficult 
to determine the
nature of the underlying compact objects. In particular, 
identification of systems
containing black holes is currently considered robust only if a dynamical mass 
is obtained. We explore a model-independent means of
identifying the central bodies --- neutron stars or 
black holes --- of accreting 
binary systems. We find four categories of 
object (classic black holes, GRS1915-like black holes, pulsars,
and non-pulsing neutron stars) occupy distinct regions in a 3-dimensional
colour-colour-intensity (CCI) diagram. Assuming that this 
clustering effect is due 
to intrinsic properties of the sources (such as mass accretion rate, binary
separation, mass ratio, magnetic field strength, etc.), 
we suggest possible
physical effects that drive each object to its specific 
location in the CCI phase space.
We also suggest a surface in this space which separates 
systems that produce jets
from those which do not, and demonstrate the use of 
CCI for identifying X-ray pulsars where a period has not been established.
This method can also be used to study sub-clustering within 
a category and may prove useful for other classes of objects, 
such as cataclysmic variables and active galactic nuclei.

\end{abstract}


\section{Introduction}

X-ray binaries (XRBs) consisting of a normal star orbiting a compact
object
owe their prominence to
one of the most efficient energy release mechanisms known:
accretion onto a compact object.  The energy
produced through accretion is released over essentially the entire
electromagnetic
spectrum, with each part of the spectrum revealing information,
often time-variable,
characteristic of particular segments of the system.

XRBs are classified into
sub-categories based on the type of the compact object
(neutron star [NS] or black hole [BH]), the
mass of the companion star (less than [LMXB] or
greater than [HMXB] 1 solar mass),
and a wide range of spectral
and temporal behaviors, including luminosity (high or low),
the presence (pulsars) or absence
of pulses (non-pulsing), jets (microquasars), outbursts
(bursters and transients), etc. (Table 1).

In spite of the diverse behavior displayed by the various categories
of sources there are
only a few accepted methods to identify the underlying compact
 object from that behavior.
Regular pulsations identify compact objects as NSs
with strong magnetic fields, yet not all NSs with
strong magnetic fields will be seen to pulse, as the orientation of
the pulsar beam and spin axis may not be favorable in some cases.
While spectral and timing analysis can hint that an object is a BH,
only a dynamical measurement of mass will be convincing.
In the four decades that they have been studied, the dynamical
technique has
unambigously identifed only 20 systems as containing BHs
with another 20 considered to be BH candidates
(BHC; Remillard \& McClintock 2006; hereafter RM06)
out of the several hundreds of
XRBs known to date (Liu, van Paradijs, \& van den Heuvel 2000; 2001,
hereafter Lvv00 and Lvv01).  Many sources remain unclassified.
Individual sources are
often found to have several different spectral states and variability
modes.
For example, Belloni~\etal~(2000) identify 12
modes of variability in the
microquasar GRS1915+105, Koljonen~\etal~(2010) identify six
states in the Wolf-Rayet system Cyg X-3.

Color-colour (CC) and colour-intensity (CI) diagrams have long been
used to classify X-ray binary types.  For example,
NS systems with low-mass companions that do
not pulse are often sub-divided by the shapes (Z and atoll)
that they trace out in X-ray CC diagrams (Hasinger \& van der Klis 1989;
Wijnands~\etal~1998).
Fender, Belloni, \& Gallo (2005) used a CI diagram to define a empirical model
for coupling between accretion and jet-production in Galactic BH 
binaries.

In some ways CI diagrams are X-ray counterparts
to the optical ``HR" or luminosity-temperature diagram which gains
its versatility from the expected power-law isoradius track given
by the Stefan-Boltzman Law.  The HR diagram works for normal stars
because there is a direct correlation between luminosity and temperature;
however, the luminosity of XRBs depends on the mass accretion rate
onto the compact object.
In 2010 Homan~\etal~used both CC and CI diagrams to conclude that Z and atoll
sources in fact differ only in mass-accretion rate.
It occurred to us that one can consider CC and CI plots to be 2D projections of a
3-dimensional CCI diagram.
We find that the XRBs observed by the All Sky Monitor (ASM; Levine et al. 1996) 
on the Rossi X-ray Timing Explorer (RXTE) can be separated into four categories 
that are uniquely identified by locus in a 3D colour-colour-intensity (CCI) diagram.
The various sub-categories cluster in a way that lets us pinpoint
them: we find that the low magnetic field NS systems (Z, atoll, and
bursters) carve out different regions than pulsars with
stronger magnetic field, and that BHs cluster in distinct regions from NSs.

This is the first in a series of papers in which we explore the
possibility of using CCI diagrams as
a means for distinguishing between
the various sub-categories of X-ray binaries.
In Section 2 we present our method for constructing CCI diagrams.  In
Section 3 we suggest possible physical interpretations of the CCI space.
In Section 4 we conclude that CCI diagrams can provide
a powerful method for classifying XRBs and note some
possibilities
for future study.

\begin{table*}
 \centering
 \begin{minipage}{140mm}
  \caption{XRB sources examined in this paper} 
\begin{tabular}{llll}
\hline
\hline
\multicolumn{4}{l}{\bf Binary Systems containing black holes (15)}\\
\multicolumn{4}{l}{\it Dynamically well determined black holes with high mass companions (HMBH; 3):}\\
\multicolumn{4}{l}{Cyg X-1; LMC X-1; LMC X-3}\\
\multicolumn{4}{l}{\it Dynamically well determined black holes with low mass companions (LMBH; 7):}\\
\multicolumn{4}{l}{J1118+480;
J1550-564;
J1650-500;
J1655-40;
GX339-4;
J1859+226;
GRS1915+105}\\
\multicolumn{4}{l}{\it Black hole candidates (BHC; 5):} \\
\multicolumn{4}{l}{J1630-472; J1739-278; J1743-322; J1758-258; J1957+115}\\
\multicolumn{4}{l}{\bf Binary Systems containing neutron stars (29)}\\
\multicolumn{4}{l}{\it Neutron star systems with high mass companions (HMXB;10):}\\
&\multicolumn{3}{l}{Pulsars (10):
J0352+309;
J1538-522;
J1901+03;
J1947+300;
J2030+375;
Cen X-3;
GX 301-2;}\\
\multicolumn{4}{l}{
Her X-1;
SMC X-1;
Vela X-1}\\
&\multicolumn{3}{l}{Non-pulsing (0; no bright examples available)}\\
\multicolumn{4}{l}{\it Neutron star systems with low mass companions (LMXB; 18):}\\
&\multicolumn{3}{l}{Pulsars (0; no bright examples available)}\\
&\multicolumn{3}{l}{Non-pulsing (19):}\\
&&\multicolumn{2}{l}{Z-sources (5):}\\
\multicolumn{4}{l}{Sco X-1;
Cyg X-2;
GX 5-1;
GX 17+2;
GX 349+2}\\
&&\multicolumn{2}{l}{Atoll sources (7):}\\
\multicolumn{4}{l}{1556-605;
1636-53;
1705-44;
1735-44;
GX 13+1;
GX 9+1;
GX 9+9}\\
&&Bursters (7):&\\
\multicolumn{4}{l}{0614+091;
1254-69; 1608-522; 1636-53; 1916-053;
Aql X-1; Ser X-1}\\
\multicolumn{4}{l}{\bf Unclassified or controversial systems (4):}\\
\multicolumn{4}{l}{Cyg X-3; Circ X-1; GX 3+1;
4U1700-37}\\
\hline
\hline
\end{tabular}
\end{minipage}
\end{table*}

\section[]{Data Analysis: Generating CCI Diagrams}

\begin{figure*}
  \vspace*{174pt}
\includegraphics[width=6.2in,angle=0]{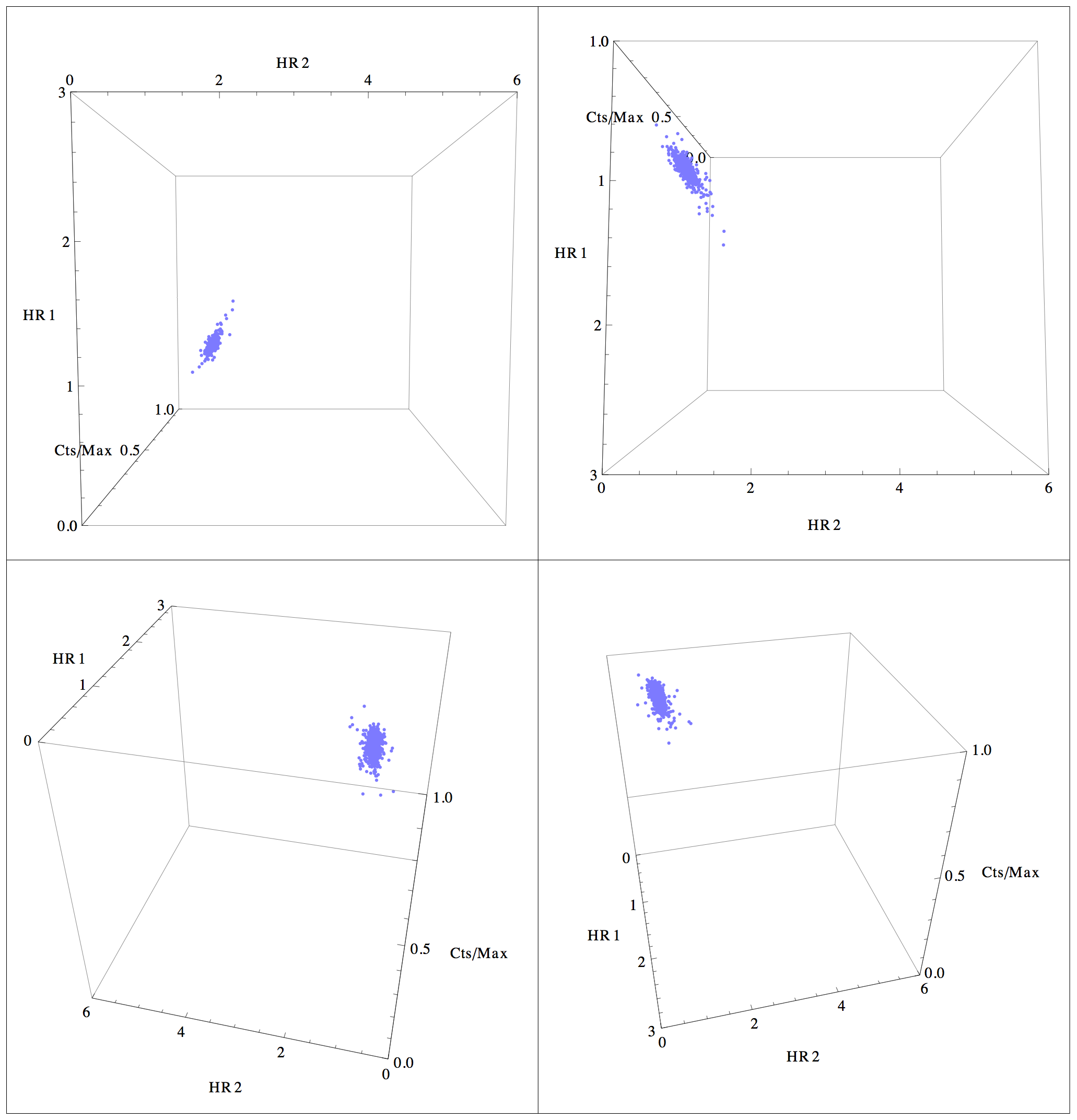}
\caption{
Four views of a CCI diagram of the first thirteen years of the RXTE/ASM observations of the
Crab. Only five sigma detections are plotted.
The ellipsoid center and radii match that for any two-year interval within the first thirteen
(see Table 2).
}
\label{crab13}
\end{figure*}

\begin{figure*}
  \vspace*{174pt}
\includegraphics[width=6.2in,angle=0]{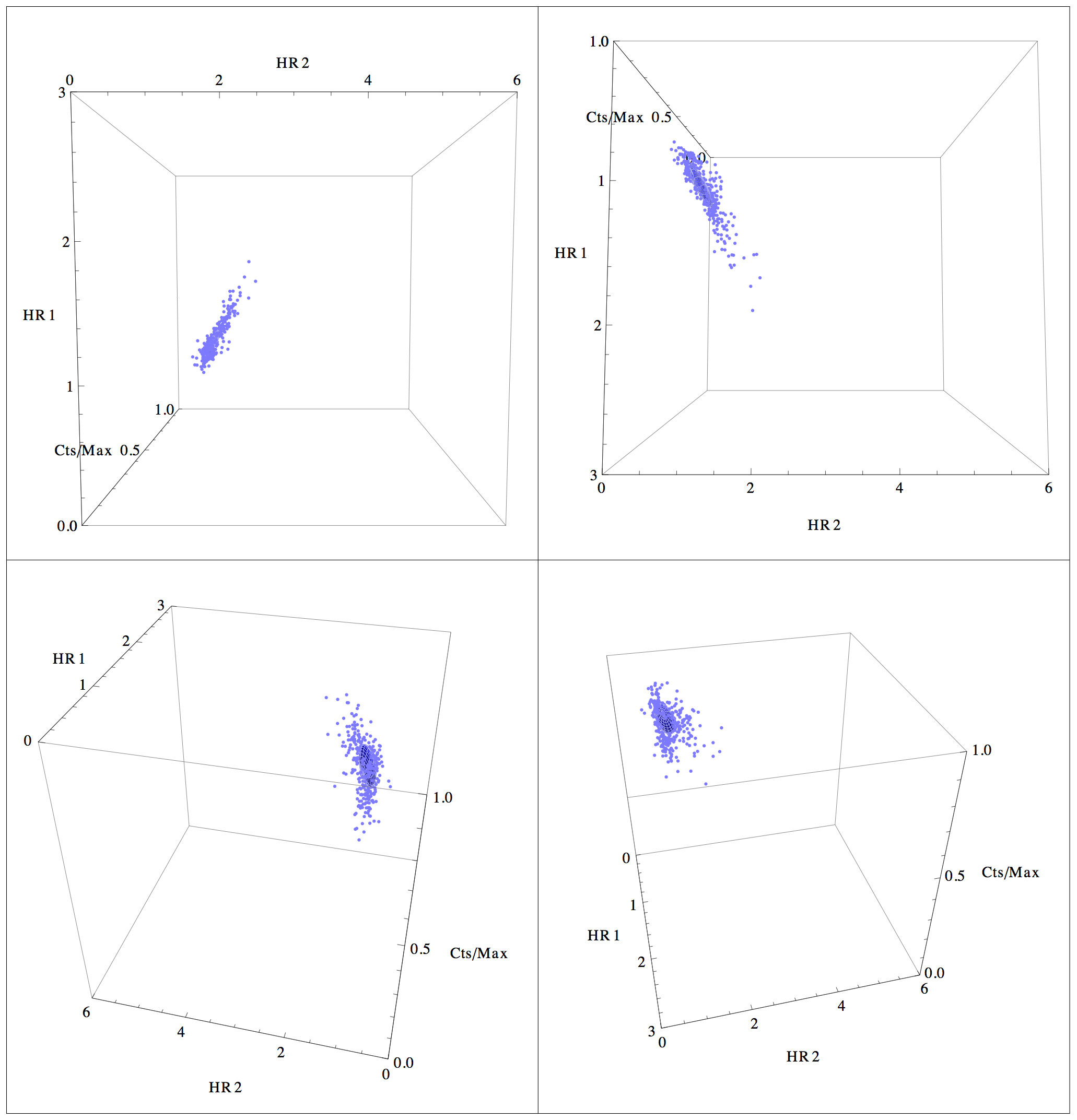}
\caption{
Four views of a CCI diagram of the last two years of the RXTE/ASM observations of the
Crab. Only five sigma detections are plotted. The ellipsoid center and radii are significantly
different from the first 13 years (see Table 2).
}
\label{crablast2}
\end{figure*}

\begin{figure*}
  \vspace*{174pt}
\includegraphics[width=6.2in,angle=0]{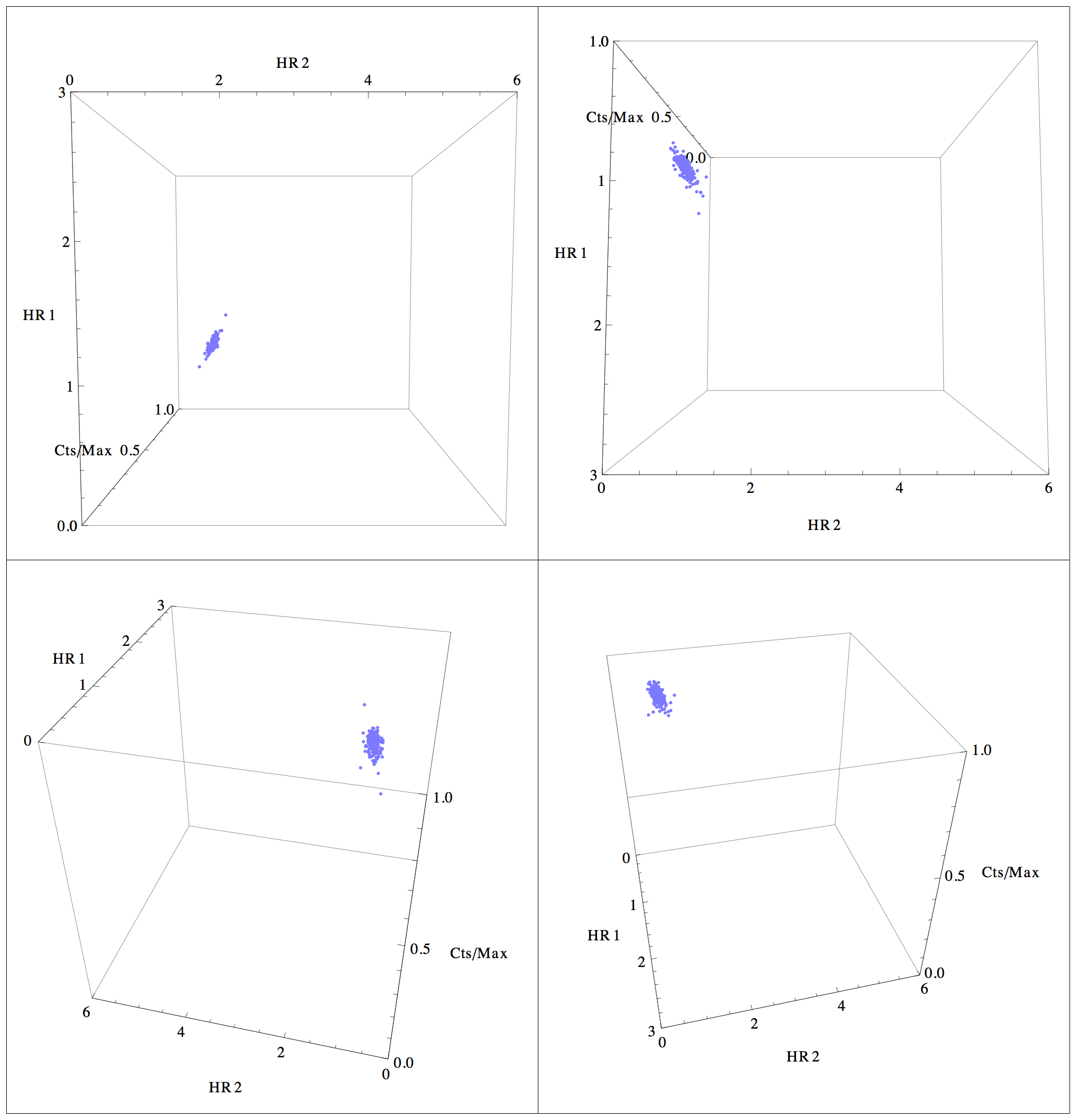}
\caption{
Four views of a CCI diagram of the 3rd and 4th years of the 
RXTE/ASM observations of the
Crab. Only five sigma detections are plotted.
The ellipsoid center and radii match that for any two-year interval within the first thirteen
as well as for the first 13 years combined (see Table 2).
}
\label{crab2ndtwo}
\end{figure*}

We use data accumulated during
the ASM lifetime
and
provided by courtesy of the MIT ASM/RXTE
team.\footnote{http://xte.mit.edu/ASM\_lc.html}
We were informed by the ASM team (A. Levine and R. Remillard, personal
communication) that there were gain changes in the instrument for
the last two years of observations. In order to test for any
long term trends in the instrument we
first divided all the data into intervals of 2 years.  We find that
there is no significant change in the placement of the points
(as determined by centroid location) over the first 6 intervals.
There is a change
in the last
two years.
Using the Crab as a test case, we find that for each of the first six 
two year intervals, as well as for the first 13 years combined,  
we compute exactly the same ellipsoid
(see Table 2). The last two-year interval
had significantly different values for the ellipsoid center and size.   
To illustrate this we show three CCI figures of the Crab: the first 
13 years as a whole (Fig. \ref{crab13}); using data from only the last two 
years (Fig. \ref{crablast2}); and a single two-year interval (Fig. \ref{crab2ndtwo}).
We note that in addition to having different ellipsoid values, the last two years
also show significantly more scatter (as shown by the size of the radii 
listed in Table 2).  We are thus confident that the first thirteen
years of ASM data show no discernable effects that can be
attributed to gain changes, and we therefore use only the first 
thirteen years as recommended
by the ASM team.

The
background of the ASM/RXTE A band
(1.3-3.0keV) can be subject to contamination by Solar UV radiation
(A. Levine and R. Remillard, personal communication).
This excess is flagged in the dwell-by-dwell data provided by the ASM
team.  By eliminating data points were this flag (column 13 of the 
dwell-by-dwell data file) exceeds 20 counts this
problem can be avoided.  There are six-eight dwells per day or
well over 32,000 points for the 15 years; the number of points out of the 
dwells in which the flag exceeds
20 counts ranges from 150-300.
We consider daily averages for our CCI diagrams and for the bright 
sources where there are over 4500 detections over 13 years, 
the possibility of 300 data points being corrupt is a 
less than 10\% effect.  For weak sources were there may be
less than 1000 detected points this can be a significant effect.
In this paper
we restrict ourselves to strong sources (average
counts greater than 2 ASM cts/sec): this removes from our sample
76 of the 122 XRBs for which we have extracted ASM light curves.
\footnote{Funding for extraction of data for all sources excluding 
points contaminated
by Solar UV is being sought; this will enable
us to use the weaker sources.}

We are using one-day averages which for 13 years corresponds to about 
4500 points
for each source.
We use the energy bands
provided by the ASM/RXTE team
 (A:1.3-3.0 keV; B:3.0-5.0keV; C:5.0-12.2 keV) to define
soft colour (HR1) as B/A and hard colour (HR2) as C/A.
Summed counts (1.3-12.2keV) are shown on the intensity axis,
scaled from 0-1 for each source.
We only use points where the signal-to-noise of the ASM/RXTE
in the summed intensity is greater than or equal to 5$\sigma$.

We have collected the sources in groups using the
classification and nomenclature from Lvv00 and Lvv01
for NS systems and RM06 for BH systems (Table 1).
We use the two hardness ratios and the summed intensity
to construct 3-D ``CCI" plots for each group.
For each category of source we compute the centroid
of all points as listed in Table 2.
Using the Mathematica\footnote{http://www.wolfram.com/mathematica/}
 program EllipsoidQuantile we compute
an ellipsoid around the
centroid that contain 50\% of all points while minimizing the
volume of the ellipsoid.

\begin{figure*}
  \vspace*{174pt}
\includegraphics[width=6.2in,angle=0]{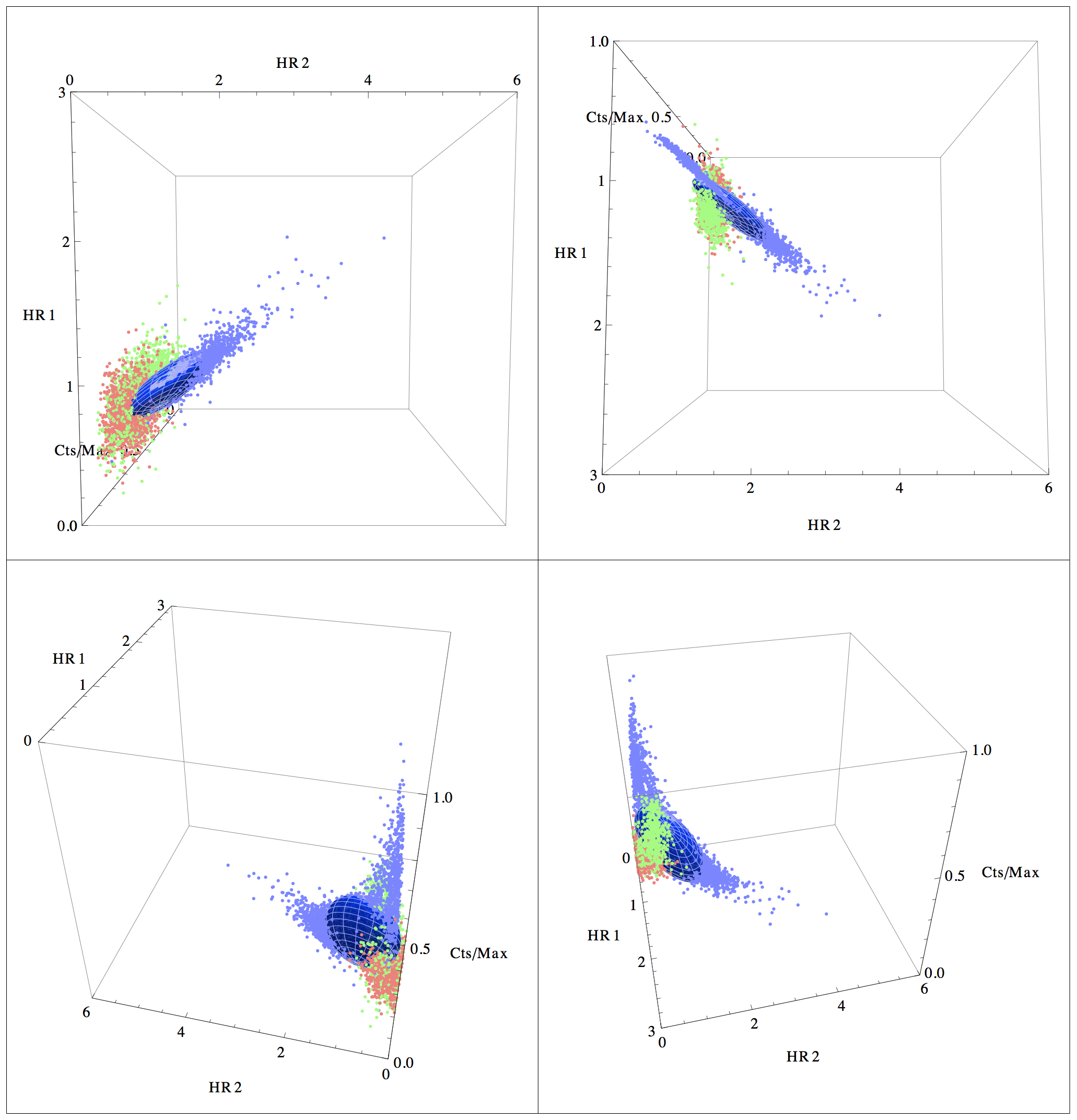}
\caption{
Four views of a CCI diagram of dynamically well-determined BHs
with high mass companions as classified by RM06.
Cyg X-1 (blue), LMC X-1 (red), LMC X-3 (green).
A single ellipsoid (blue) has been fit to
encompass 50\% of all
the points, centered on the centroid of all points.
See Table 2 for values of the centroid.
}
\label{hmbhs}
\end{figure*}

\begin{figure*}
  \vspace*{174pt}
\includegraphics[width=6.2in,angle=0]{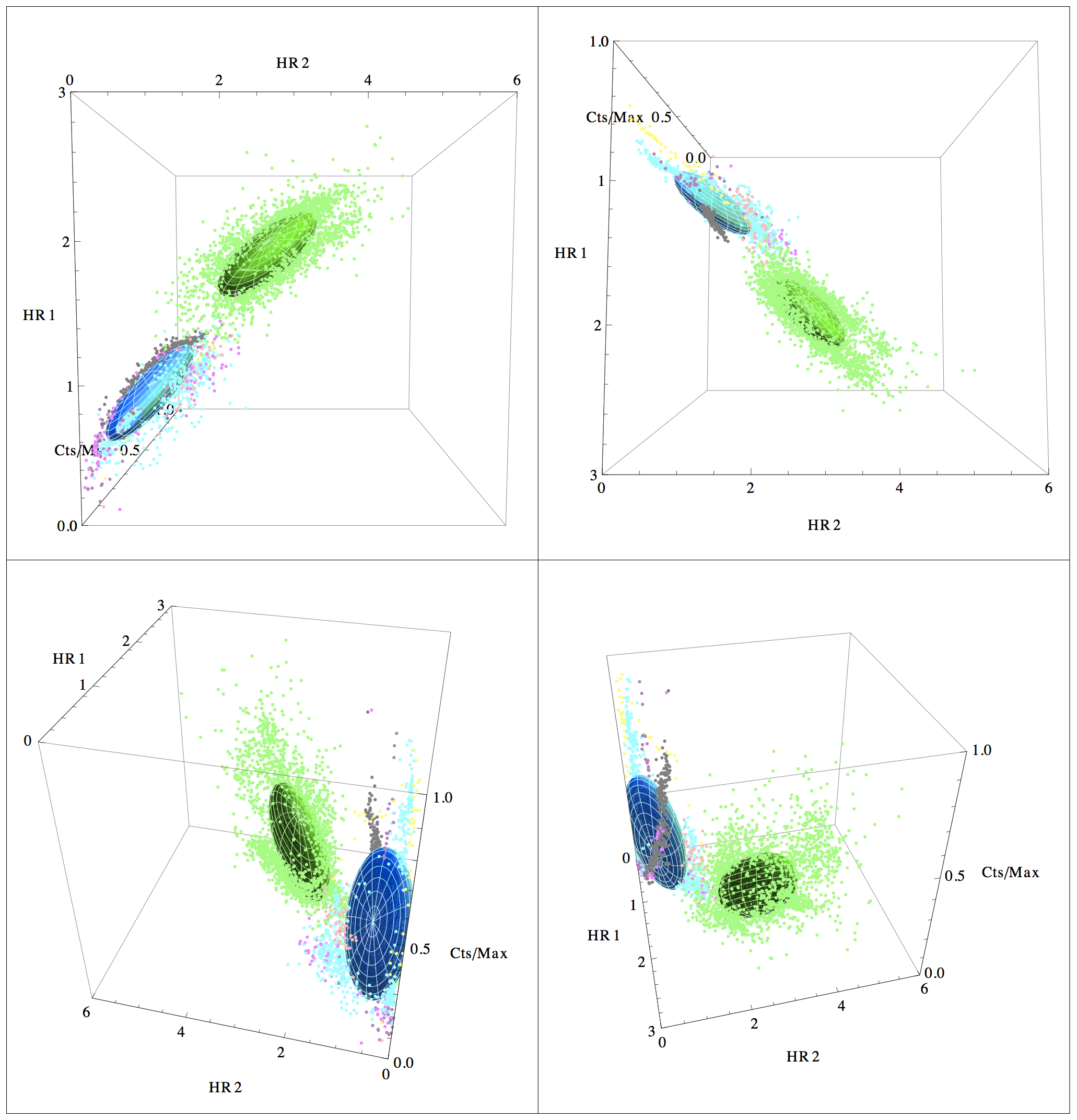}
\caption{
Four views of a CCI diagram of dynamically well-determined BHs
with low-mass companions as listed in Table 1.
1118+480 (pink),
1550-564 (magenta), 1650-500 (yellow), 1655-40 (cyan), 1859+226 (purple), GRS1915+105 (green),
GX339-4 (black).  We note that GRS 1915+105 (green) occupies a space
well separated from the rest of the LMXB BHs. 
Ellipsoids fitted to all LMBH excluding GRS1915+105
(cyan) and separately to GRS 1915+105 (green).
See Table 2 for values of the centroids.}
\label{lmbhs}
\end{figure*}

\begin{figure*}
  \vspace*{174pt}
\includegraphics[width=6.2in,angle=0]{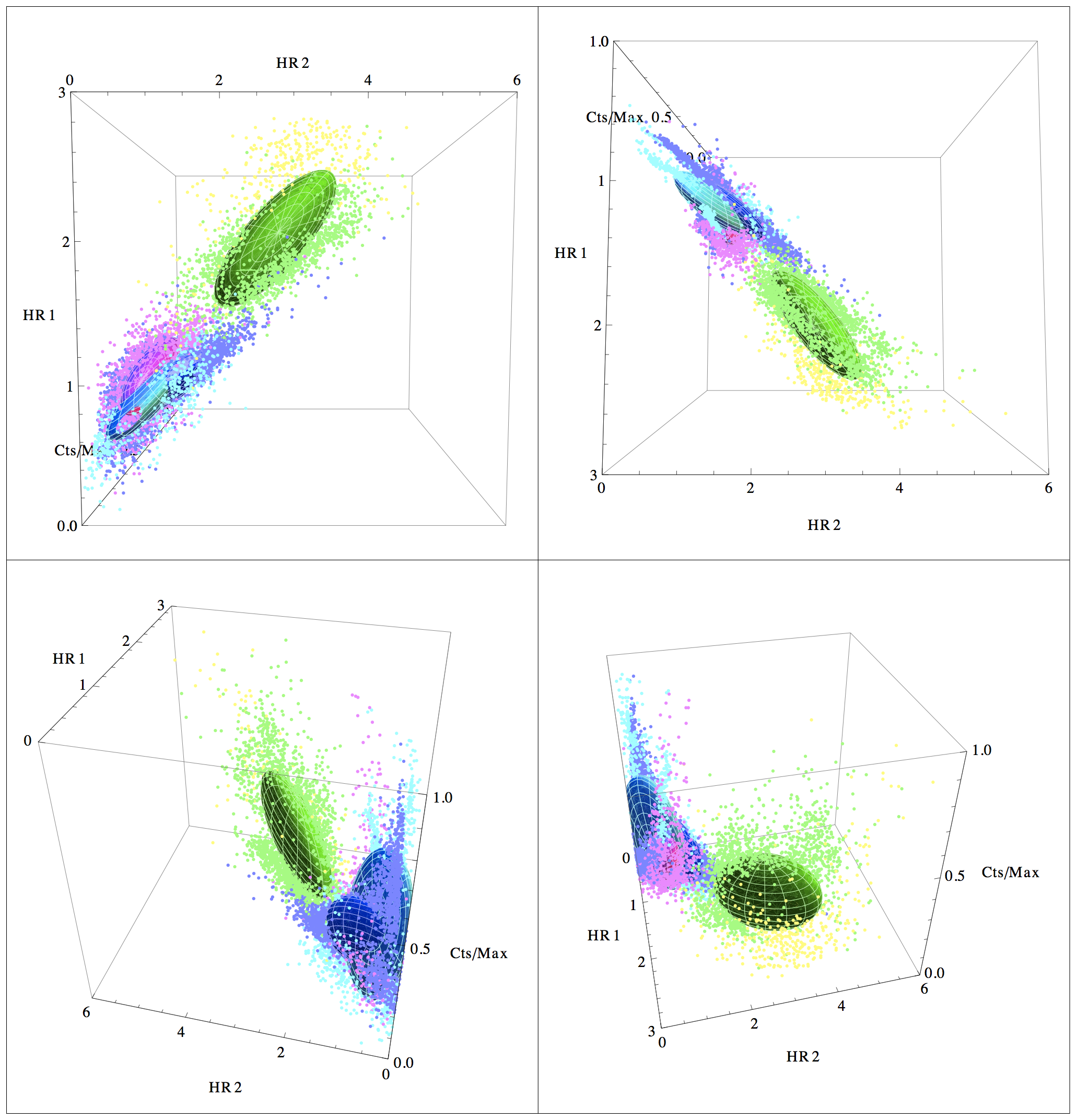}
\caption{Four views of a CCI diagram of all dynamically well determined 
BHs with massive companions in blue, with low mass companions in cyan, all BHC 
candidate systems in magenta,  
GRS 1915+105 in green and 
 J1630-475 in yellow (a single ellipsoid in green is fit to both these). 
See Table 2 for values of the centroids.
}
\label{bhbhc}
\end{figure*}

Figures \ref{hmbhs} and \ref{lmbhs} shows CCI diagrams
of all systems identified by RM06 as definitely
containing BHs.
 We
have separated systems with high-mass companions from systems
with low-mass companions.
The data of each object within a group are shown in different
colours but the ellipsoid represents all objects of a given group.
There is significant overlap between BH systems with low-mass and high-mass
companions with the
exception of GRS1915+105 (a rare microquasar that ejects
material at superluminal velocities; Mirabel \& Rodriguez 1994).
We have separated out GRS1915+105 because it occupies a space that is 
disjoint from the other BH systems. 
We then considered systems identified by RM06 as
BHCs.  In Figure \ref{bhbhc}
we depict BHs with high-mass companions in blue, BHs
with low-mass companions (with the exception of GRS1915+105) in cyan, 
BHCs (with the exception of J1630-472) in magenta.   
The ellipsoids and points for 
 BHs and BHCs (with both low-mass and high-mass companions)
overlap, whereas GRS1915+105 and J1630-472 clearly occupy
a region distinct from the others. 
\begin{table*}
 \centering
 \begin{minipage}{140mm}
  \caption{Centroids and radii}
\begin{tabular}{lll}
\hline
\hline
 Source Type  &  Center   &       radius1 radius2  radius3\\
\hline
&&\\
 Fig1 Crab13 & 0.87, 0.95, 0.95& 0.05,0.02,0.02\\
 Fig2 Crablast2&0.94, 1.05, 0.86 & 0.25,0.06,0.06\\ 
 Fig3 Crab2&0.87, 0.95, 0.95&0.05,0.02,0.02\\ 
&&\\
 Fig4 HMBHs    & 0.83, 0.87, 0.25 &  0.84,0.18,0.15\\
&&\\
 Fig5 LMBHs; no GRS1915+105&
                0.79,0.56,0.30 & 0.72,0.32,0.19\\
 Fig5 GRS1915+105      &  1.96,2.60,0.28 &   0.86,0.25,0.16\\
&&\\
 Fig6 HMBHs     &  0.83,0.87,0.25   &   0.84,0.18,0.15\\
 Fig6 LMBHs     &  0.79,0.56,0.30   &   0.72,0.32,0.19\\
 Fig6 BHCs&0.95,0.68,0.22& 0.41,0.23,0.13\\
 Fig6 GRS1915+105 and 1630-472&2.09,2.75,0.28  &  1.13,0.32,0.17\\
&&\\
 Fig7 HMXB pulsars &
                1.45,3.21,0.34  & 1.74,0.30,0.22\\
&&\\
 Fig8 HMXB pulsars no outbursts&1.37,3.14,0.29&2.15,0.30,0.13\\
 Fig8 HMXB pulsars with outbursts&
 1.52,3.28,0.39 &  1.26,0.33,0.22\\
&&\\
 Fig9 Bursters& 0.95,1.02,0.39&   0.44,0.24,0.15\\
&&\\
 Fig10 Atolls   &  1.31,1.61,0.51 &  0.89,0.20,0.19\\
&&\\
 Fig11 Z sources & 1.27,1.55,0.49  & 0.77,0.12,0.10\\
&&\\
 Fig12 Classic BHs     &  0.82,0.80,0.26   &   0.87,0.21,0.19\\
 Fig12 GRS1915-like BHs     &  2.09,2.75,0.28   &   1.13,0.32,0.17\\
 Fig12 Pulsars &  1.45,3.21,0.34 &  1.74,0.30,0.22\\
 Fig12 Non-pulsing NS&  1.31,1.63,0.53 &  0.84,0.17,0.15\\
&&\\
 Fig17 Bursters & 0.95,1.02,0.39 &  0.44,0.24,0.15\\
 Fig17 Atolls   & 1.31,1.61,0.51  & 0.89,0.20,0.19\\
 Fig17 Z sources & 1.27,1.55,0.49  & 0.77,0.12,0.10\\
&&\\
Fig19 BHs with resolved jets&1.25,1.53,0.33&1.37,0.24,0.20\\
Fig19 Pulsars &1.45,3.21,0.34&1.74,0.30,0.22\\
&&\\
\hline
\hline
\end{tabular}
\end{minipage}
\end{table*}

\begin{figure*}
  \vspace*{174pt}
\includegraphics[width=6.2in,angle=0]{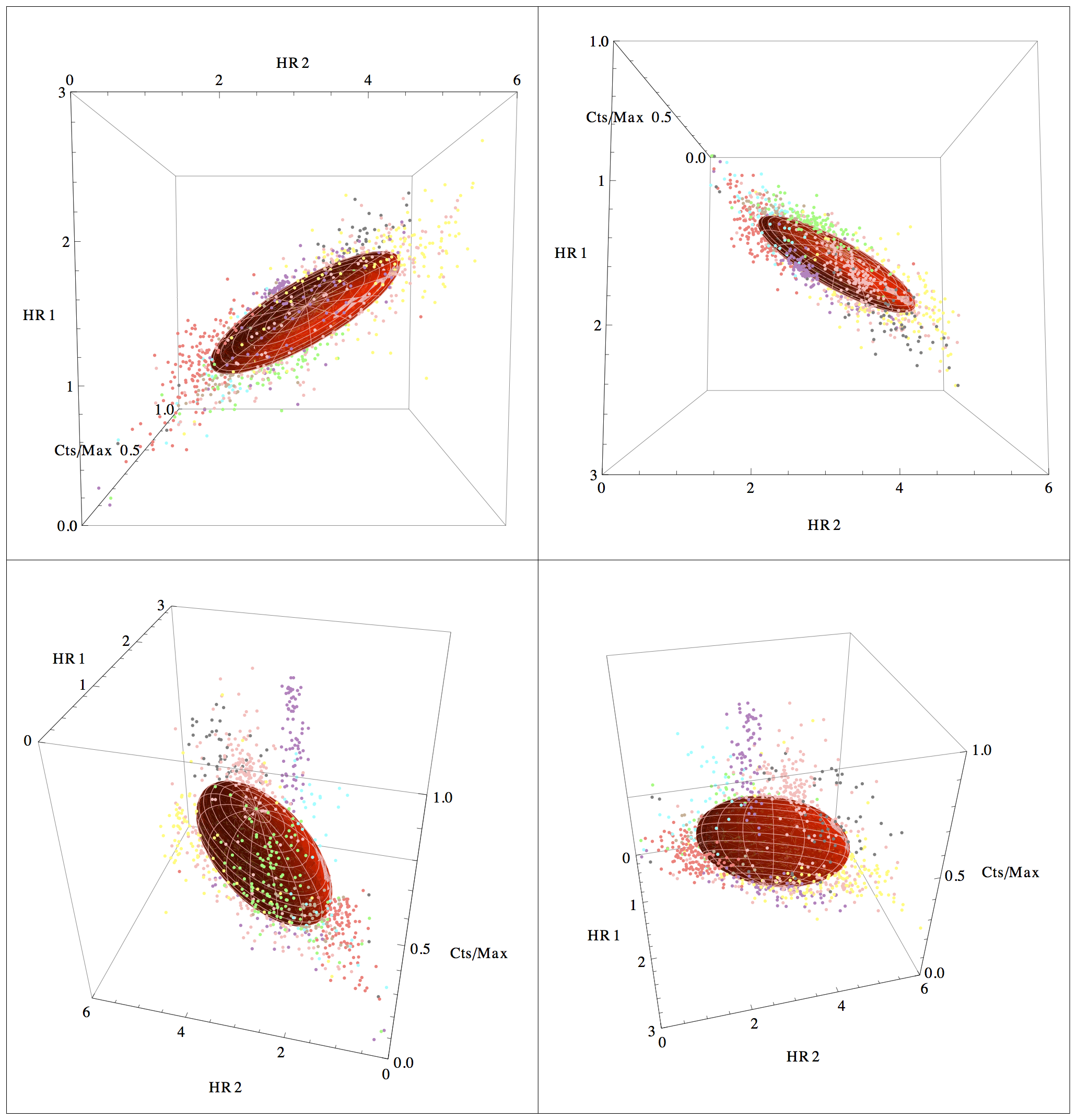}
\caption{Four views of a CCI diagram of systems with high mass
companions where the NS is a pulsar.
Different colors represent different sources.
The sources that stick up (1947+300 in cyan; 2030+375 in purple;
Cen X-3 in pink; 1901+03 in red)
are sources that
have outburst states.
A single 50\% ellipsoid (red) is fitted to all points.
See Table 2 for values of the centroid.}
\label{hmpulsars}
\end{figure*}

\begin{figure*}
  \vspace*{174pt}
\includegraphics[width=6.2in,angle=0]{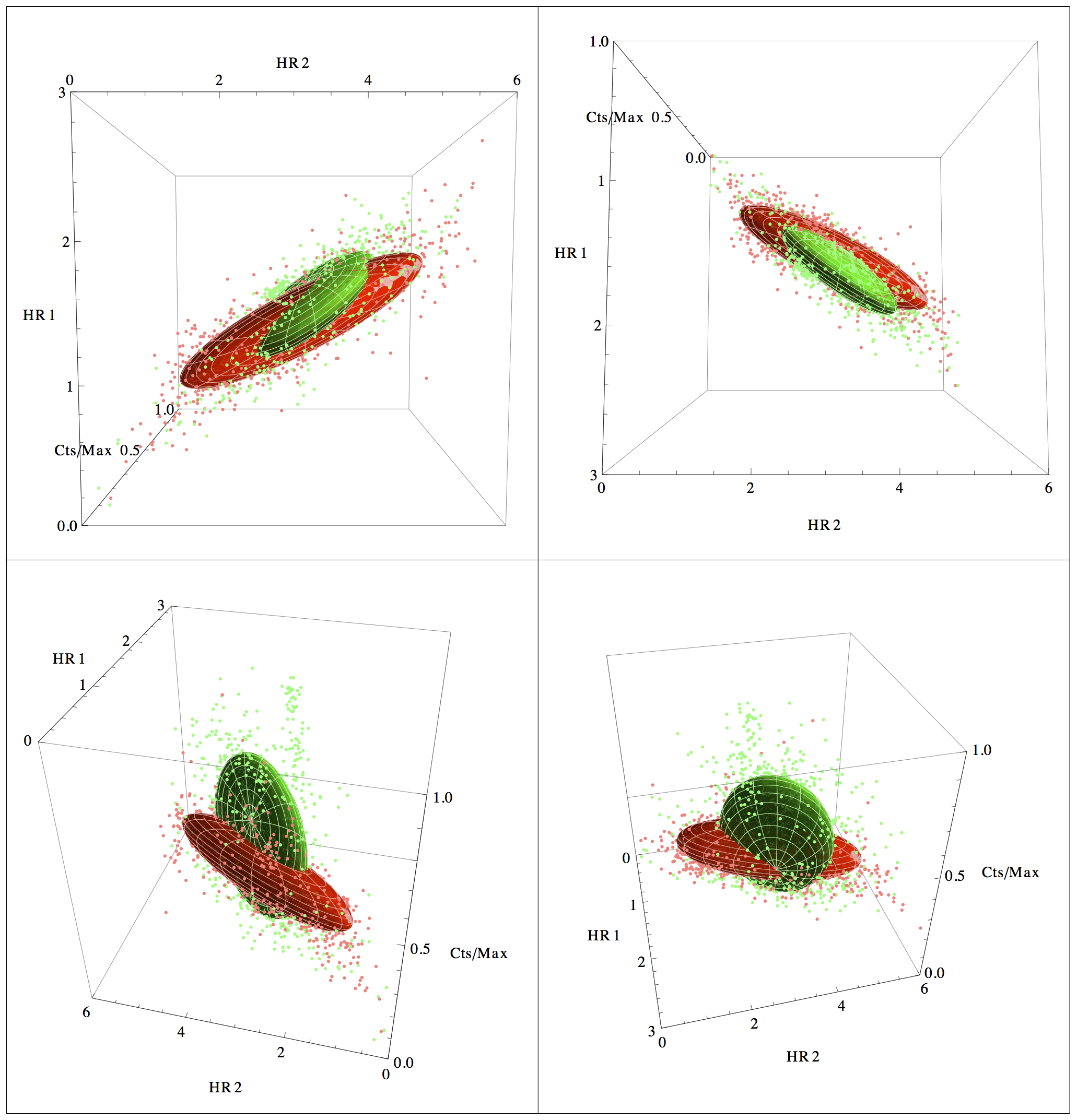}
\caption{ CCI diagrams of systems
where the NS is a pulsar.
50\% ellipsoids fitted to pulsars with outbursts (green) and
all other pulsars (red).
See Table 2 for values of the centroids.
}
\label{hburpulsars}
\end{figure*}

\begin{figure*}
  \vspace*{174pt}
\includegraphics[width=6.2in,angle=0]{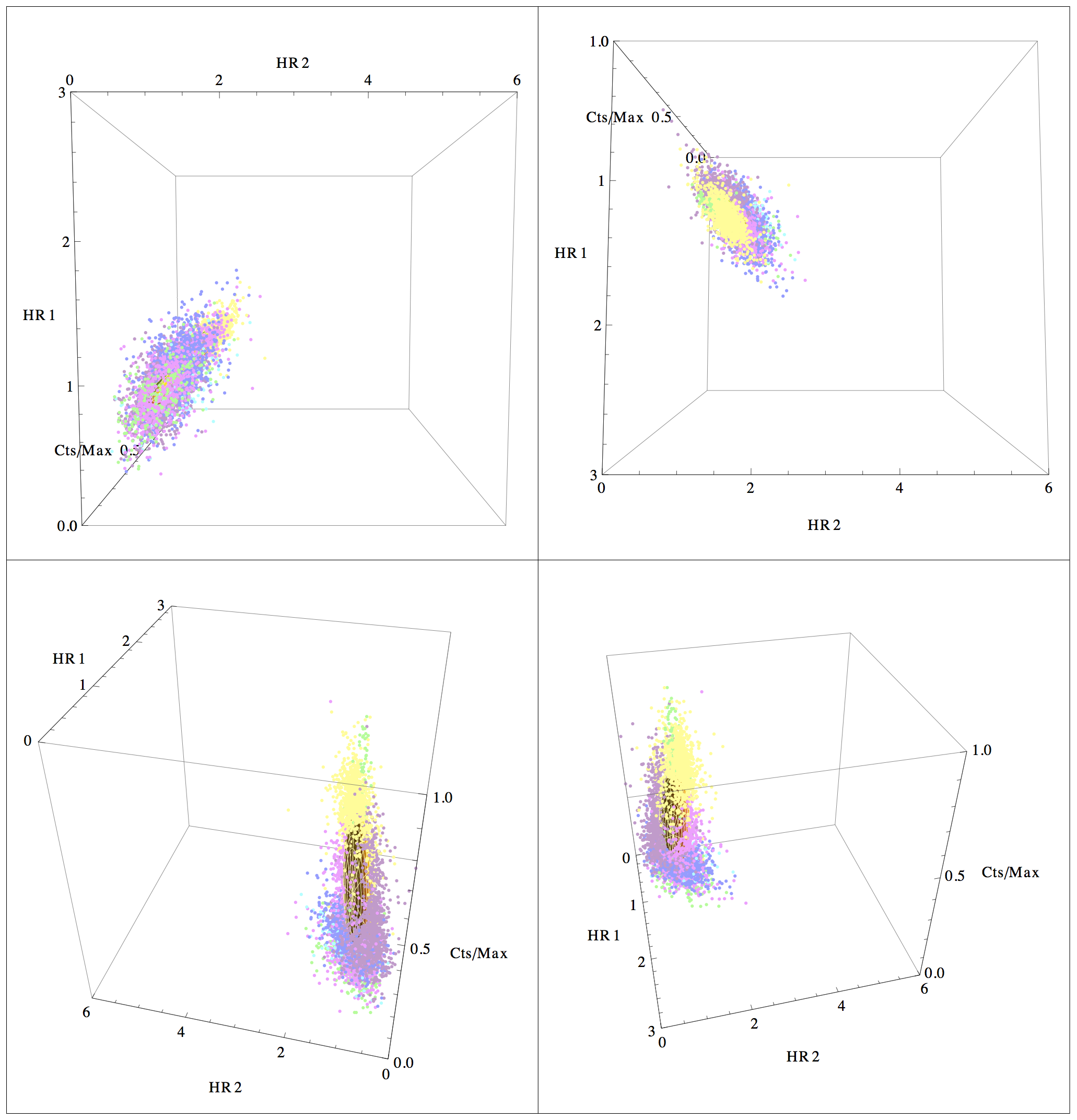}
\caption{Different colours represent different sources.
CCI of NS systems containing low-mass companions
and identified
as bursters.
50\% ellipsoid fitted to all points (red).
See Table 2 for values of the centroid.
}
\label{bursters}
\end{figure*}

\begin{figure*}
  \vspace*{174pt}
\includegraphics[width=6.2in,angle=0]{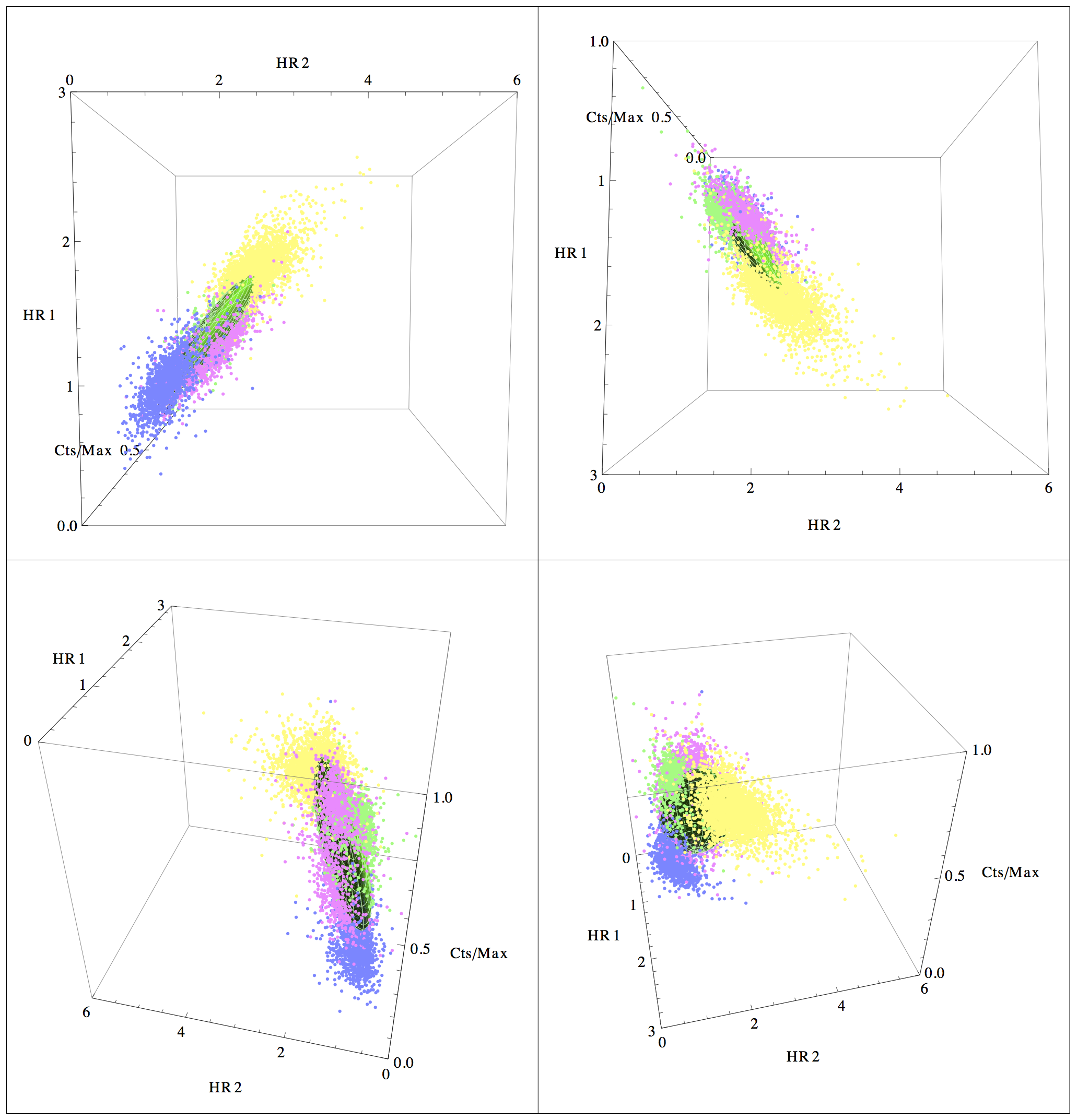}
\caption{Different colours represent different sources.
CCI of NS systems containing low-mass companions
and identified
as atoll sources. These overlap with bursters but go out to
larger HR1, HR2 values.
50\% ellipsoid fitted to all points (green).
See Table 2 for values of the centroid.
}
\label{atolls}
\end{figure*}

\begin{figure*}
  \vspace*{174pt}
\includegraphics[width=6.2in,angle=0]{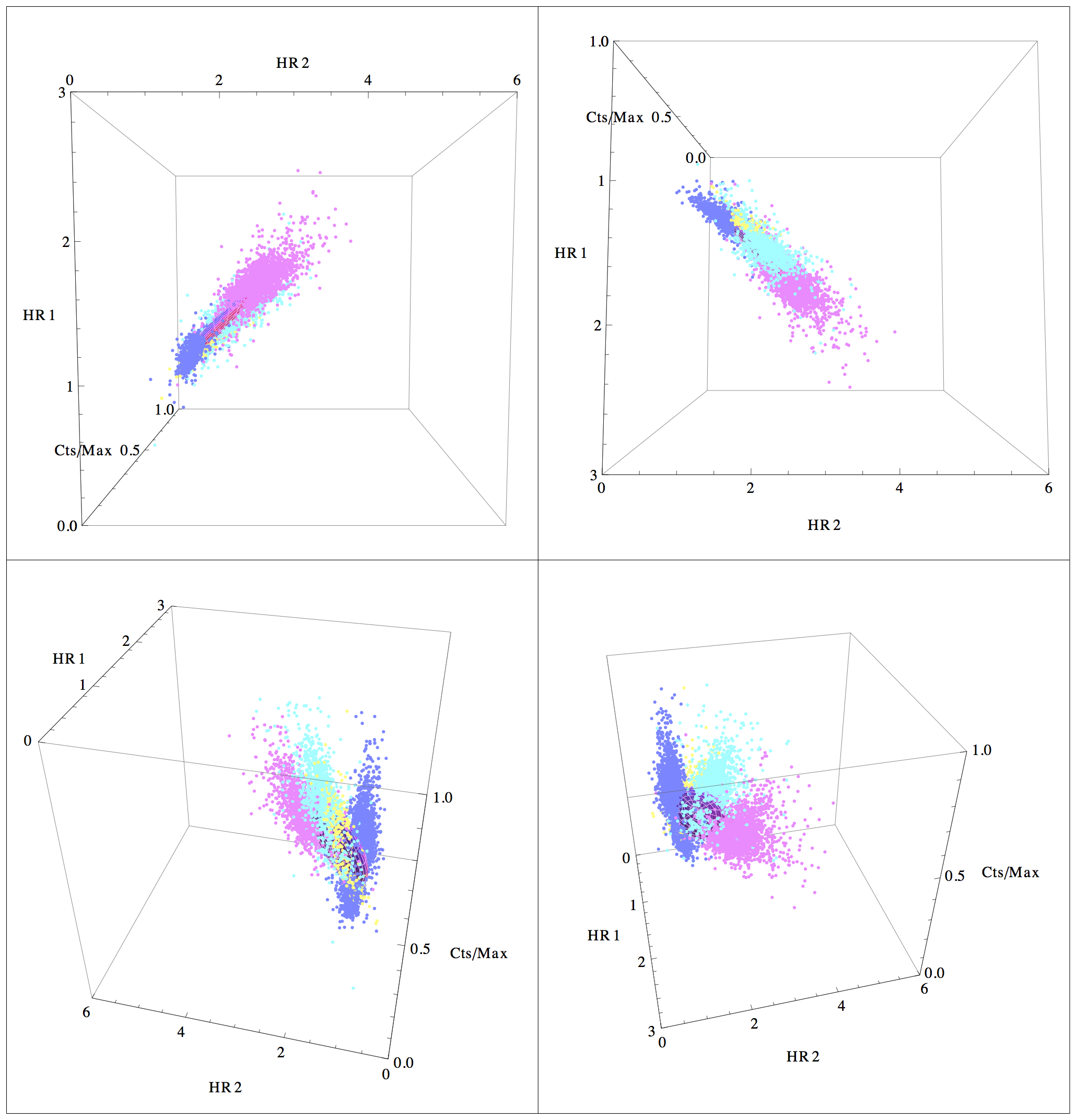}
\caption{ Different colours represent different sources.
CCI of NS systems identified as Z sources.
(Scox1 in yellow; Cyg X-2 in blue; GX17+2 in magenta; GX349+2 in cyan).
50\% ellipsoid fitted to all points (magenta).
See Table 2 for values of the centroid.
}
\label{zsources}
\end{figure*}

Figure \ref{hmpulsars} shows systems where the NS is a pulsar.
Only pulsars with high-mass companions are included as no pulsars
with low-mass companions match our selection criterion.
The sources that stick out from the bulk
of the pulsars are ones that show X-ray outbursts.
Figure \ref{hburpulsars}
shows HMXB pulsars differentiated by those which do and do not show
outbursts.
The non-pulsing NS systems (bursters, atolls, and Z-sources)
are plotted in Figures \ref{bursters}, \ref{atolls},
and
\ref{zsources}.

\begin{figure*}
  \vspace*{174pt}
\includegraphics[width=6.2in,angle=0]{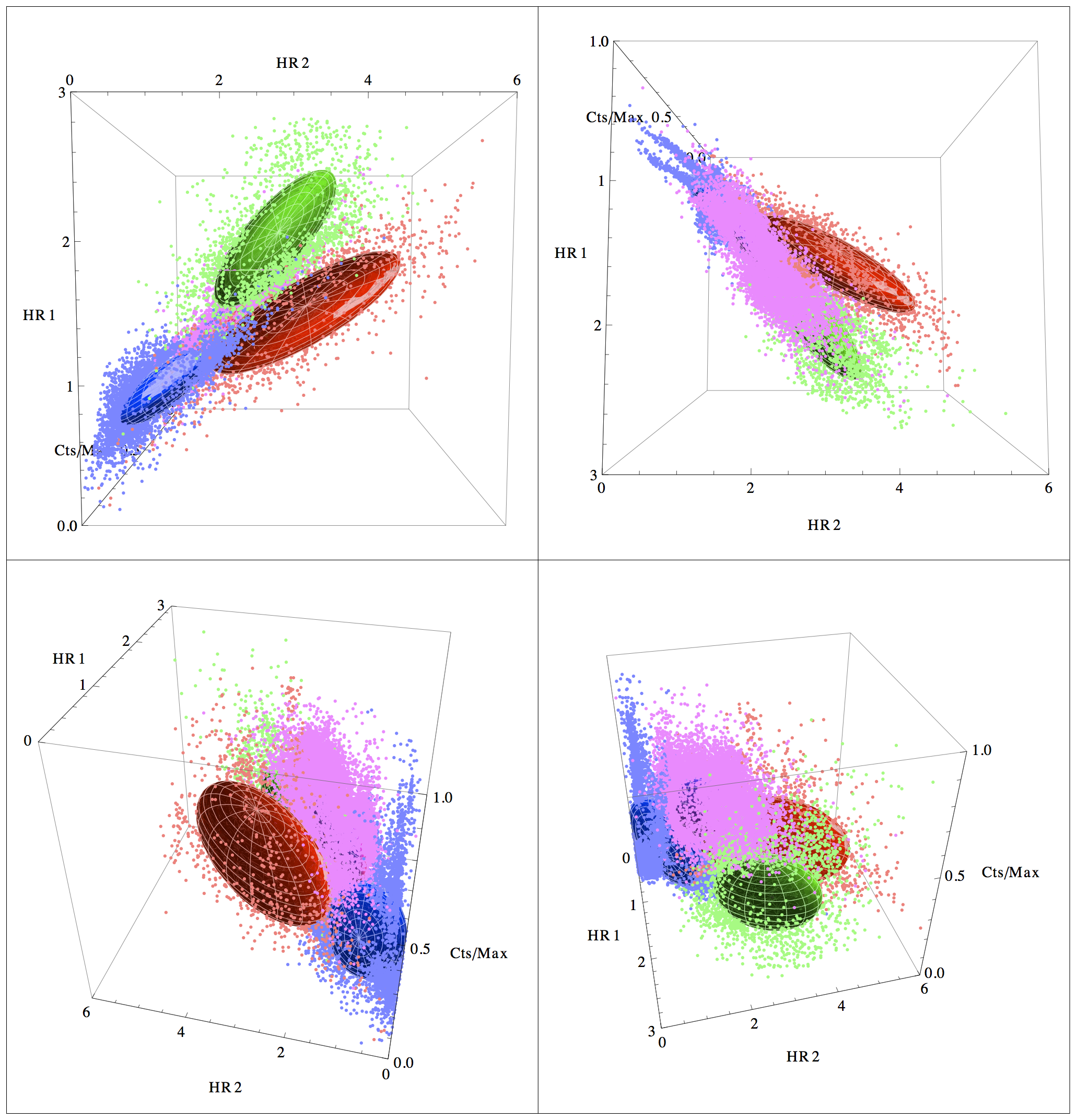}
\caption{
Four views of a CCI diagram of classic BHs 
binaries
in blue; GRS1915-like BHs in green; 
pulsars 
in red, and
non-pulsing NS systems in magenta.
ellipsoid.
It is clear that
pulsars (red), classic BHs (blue), GRS1915-like BHs(green),
and non-pulsing NS (magenta) are located
in distinct regions.
See Table 2 for values of the centroids.
}
\label{cci5sigma}
\end{figure*}

Figure \ref{cci5sigma} shows classic and GRS1915-like BHs, 
and pulsing and non-pulsing NS systems
on one plot.
Classic BHs are depicted in blue, GRS1915+105 and 
J1630-475 are depicted in green, 
NS systems that pulse are depicted in red; and NS
systems that do not pulse  are depicted in magenta.
Table 2 lists the centroid locations and length of each radius.
While some of the 2D projections in Figure \ref{cci5sigma} 
may appear to show overlap between
the different types, other projections of the same figure
 make it clear that there
is in fact no overlap; not only for the ellipsoids but indeed for any of
the points that extend beyond the ellipsoids.

\begin{figure*}
\vspace*{174pt}
\includegraphics[width=6.2in,angle=0]{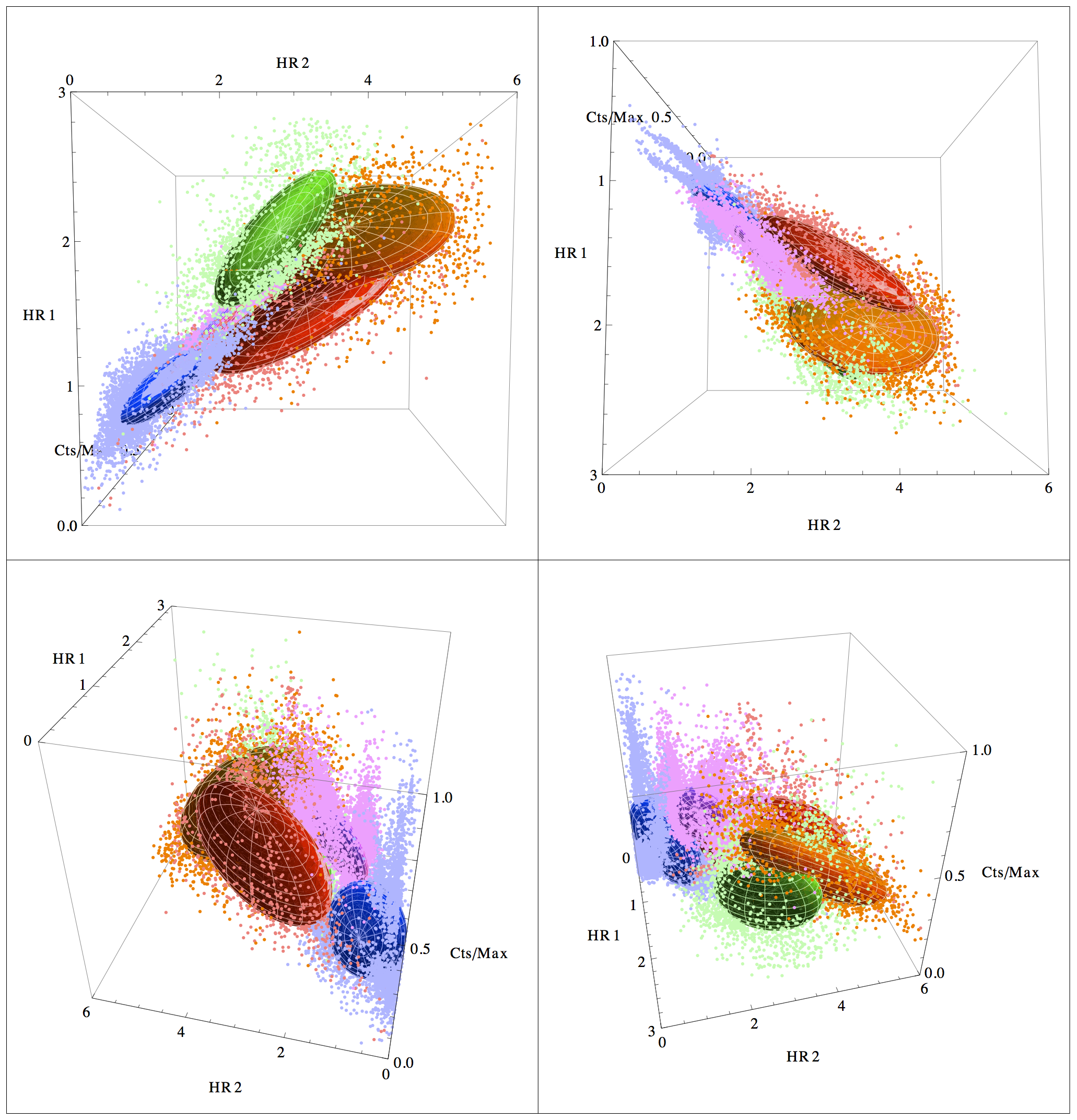}
\caption{
Comparison of unclassified sources Cyg X-3 (orange)
 with classic BH systems in blue;
GRS1915-like BH systems in green; pulsars in red; and non-pulsing
NS in magenta.
Cyg X 3 (orange) overlaps
with both GRS1915-like and pulsars. 
}
\label{cygx3comp}
\end{figure*}

\begin{figure*}
  \vspace*{174pt}
\includegraphics[width=6.2in,angle=0]{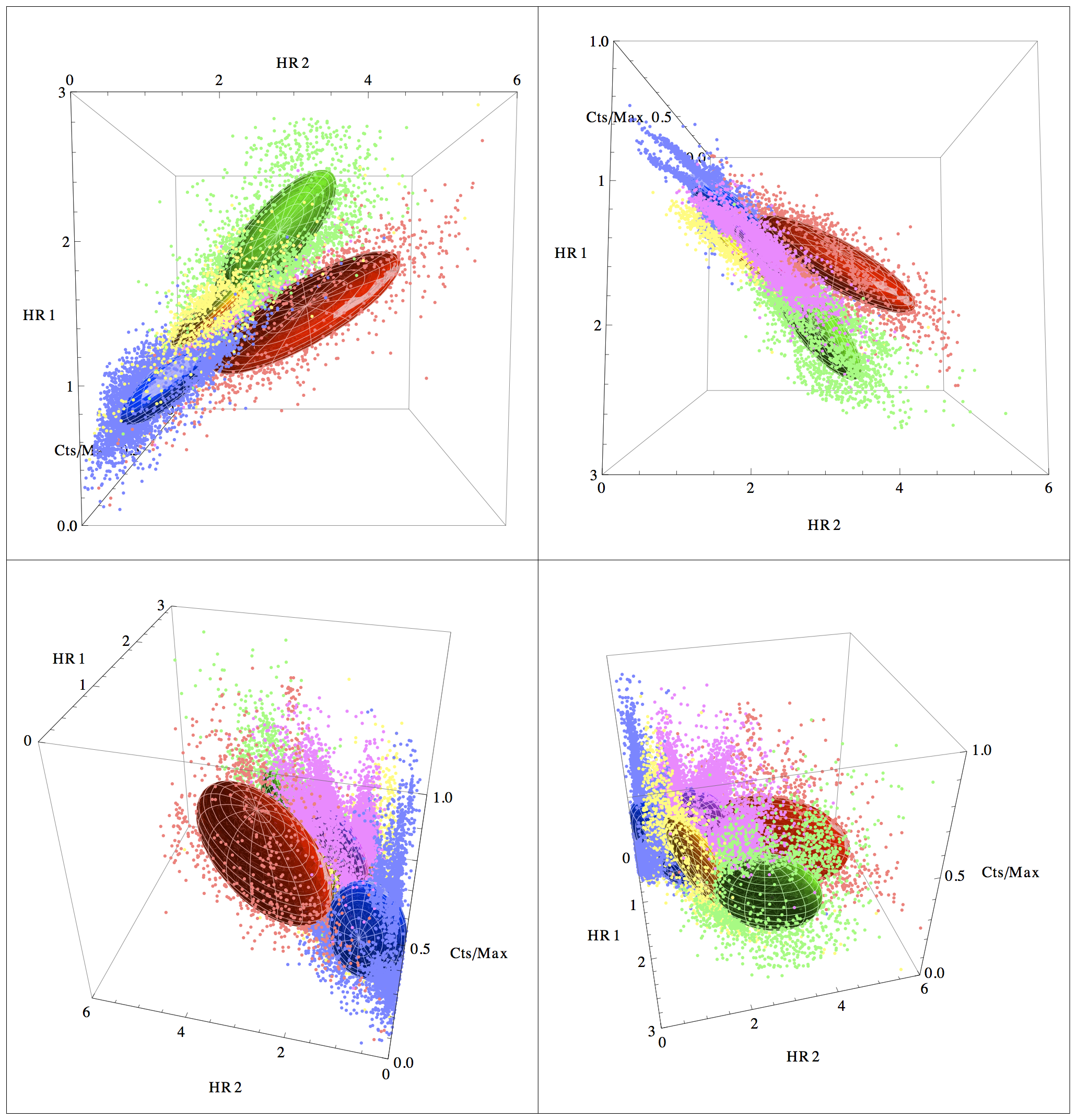}
\caption{
Comparison of unclassified sources Circ X-1 (yellow)
with classic BH systems in blue;
GRS1915-like BH systems in green; pulsars in red; and non-pulsing
NS in magenta.
Circ X-1 overlaps with non-pulsing NS.
}
\label{circx1comp}
\end{figure*}

\begin{figure*}
  \vspace*{174pt}
\includegraphics[width=6.2in,angle=0]{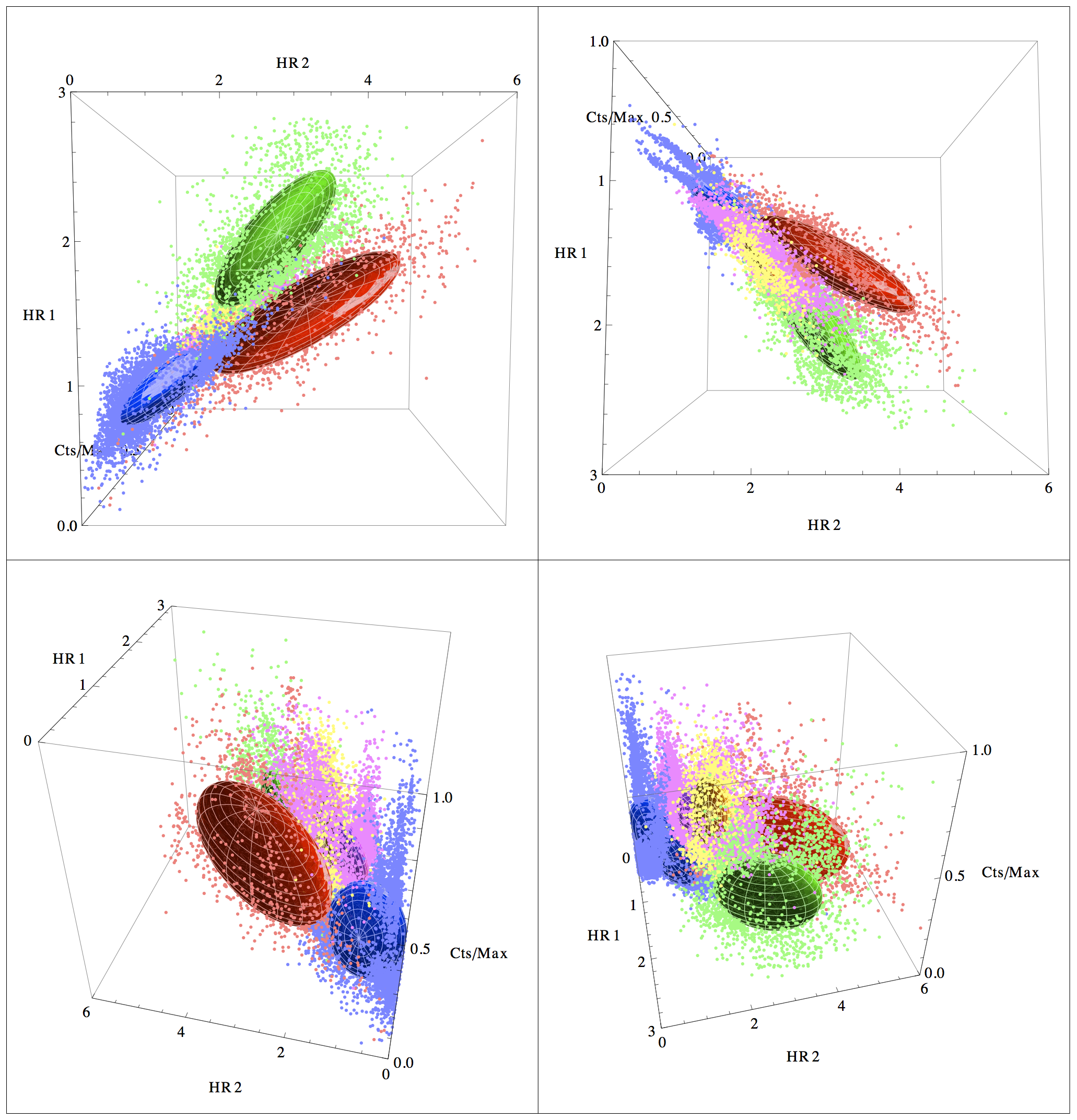}
\caption{Comparison of unclassified sources GX 3+1 (yellow)
with classic BH systems in blue;
GRS1915-like BH systems in green; pulsars in red; and non-pulsing
NS in magenta.
GX 3+1 is consistent with being a Z-source.
}
\label{gx3p1comp}
\end{figure*}

\begin{figure*}
  \vspace*{174pt}
\includegraphics[width=6.2in,angle=0]{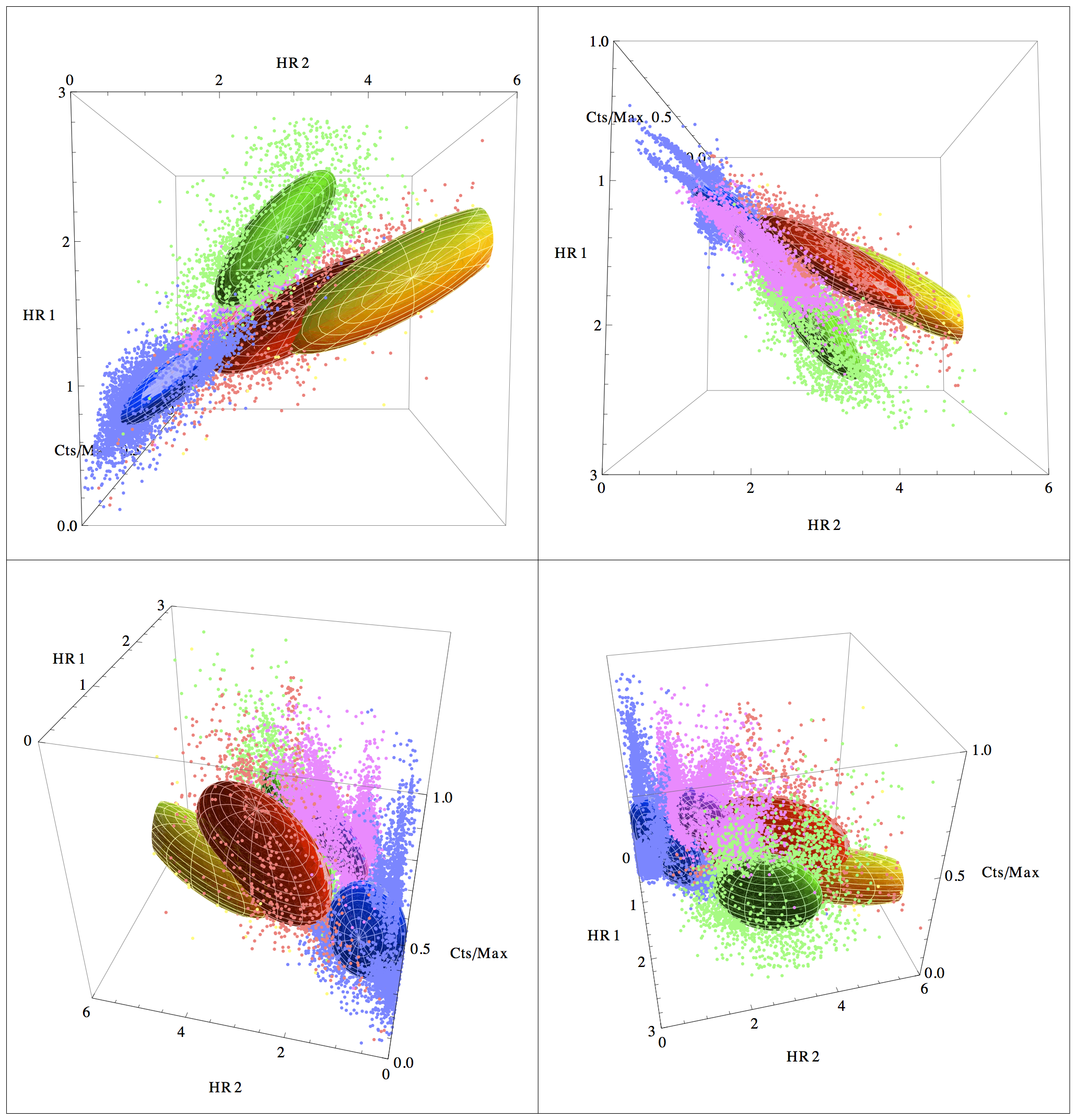}
\caption{
Comparison of unclassified source 1700-37 (yellow)
with classic BH systems in blue;
GRS1915-like BH systems in green; pulsars in red; and non-pulsing
NS in magenta.
4U1700-37 is consistent with being a pulsar.
}
\label{1700comppul}
\end{figure*}

We use the locations of classes
of objects as
pointed out above to classify the hitherto ambiguous systems
empirically as listed in Table 3.
It is immediately clear why some of these sources are difficult to classify:
they overlap categories defined hitherto.
For example,  Cyg X-3  overlaps
 with GRS1915+105 which is classified by both RM06 and
Massi \& Bernado (2008;
hereafter MB08) as
a BH, but also spends a significant amount of its time in the region
we associate sith pulsars 
(Fig. \ref{cygx3comp}).
Circinus X-1 classified by MB08 as a NS system does overlap
with non-pulsing NS 
(Fig. \ref{circx1comp}).
GX3+1 is clearly associated with non-pulsing
NS (Fig. \ref{gx3p1comp}), and
1700-37 is clearly associated with pulsars (Fig. \ref{1700comppul}).

\section{Interpreting CCI diagrams}

\begin{figure*}
  \vspace*{174pt}
\includegraphics[width=6.2in,angle=0]{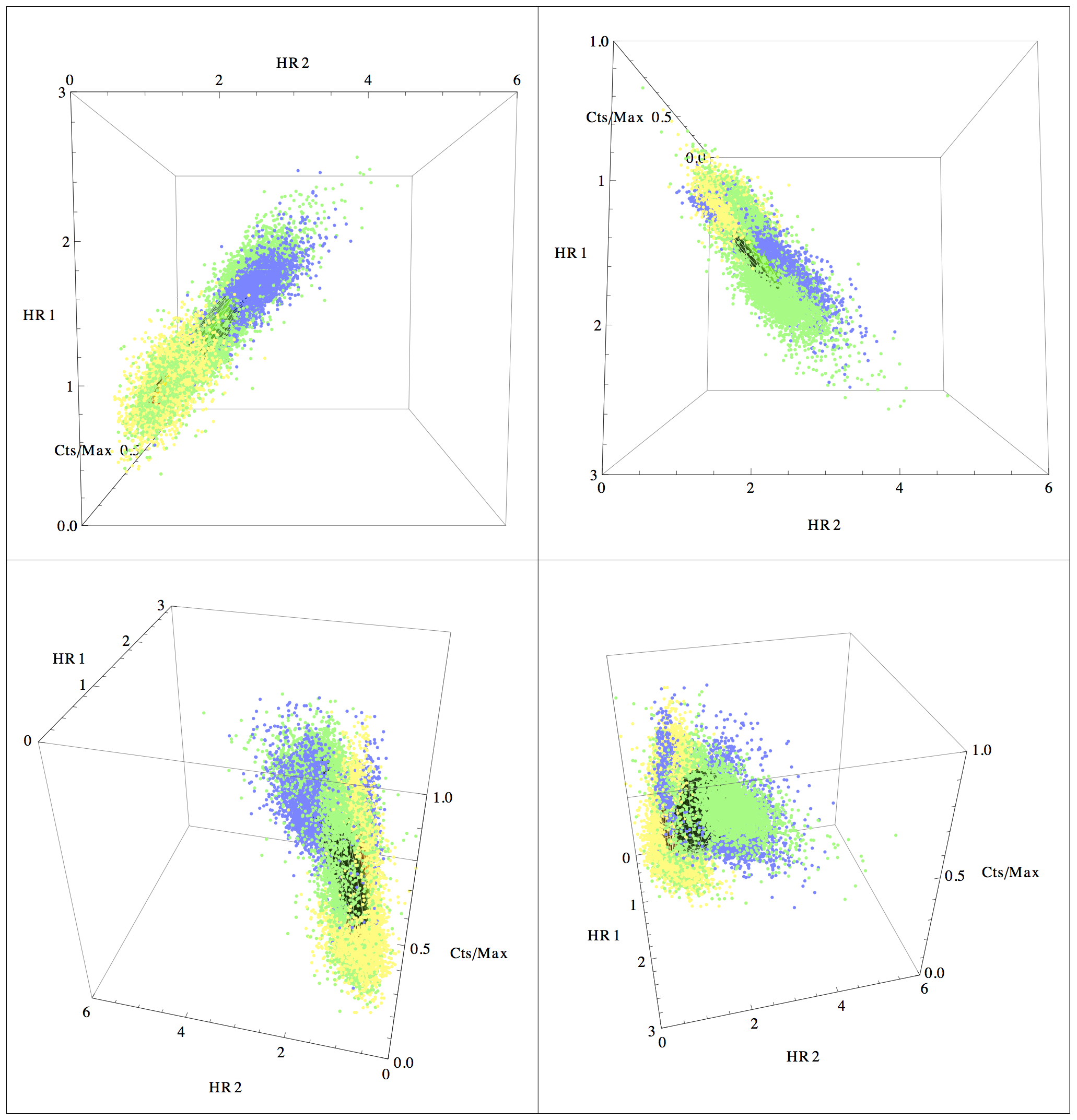}
\caption{
CCI diagram of bursters (yellow), atoll (green),
and Z-sources (cyan).
 If we accept the
premise that atolls and Z-sources are at higher mass accretion rates
than Atolls with both higher than bursters than we can infer a direction for increasing mass accretion rate.
Note that mass accretion rate does not have a simple relationship to
the X-ray flux of the systems.
50\% ellipsoids fitted to bursters (yellow);
atolls (green), Z-sources (blue).
See Table 2 for values of the centroids.
}
\label{zsatollbur}
\end{figure*}

\begin{figure*}
  \vspace*{174pt}
\includegraphics[width=6.2in,angle=0]{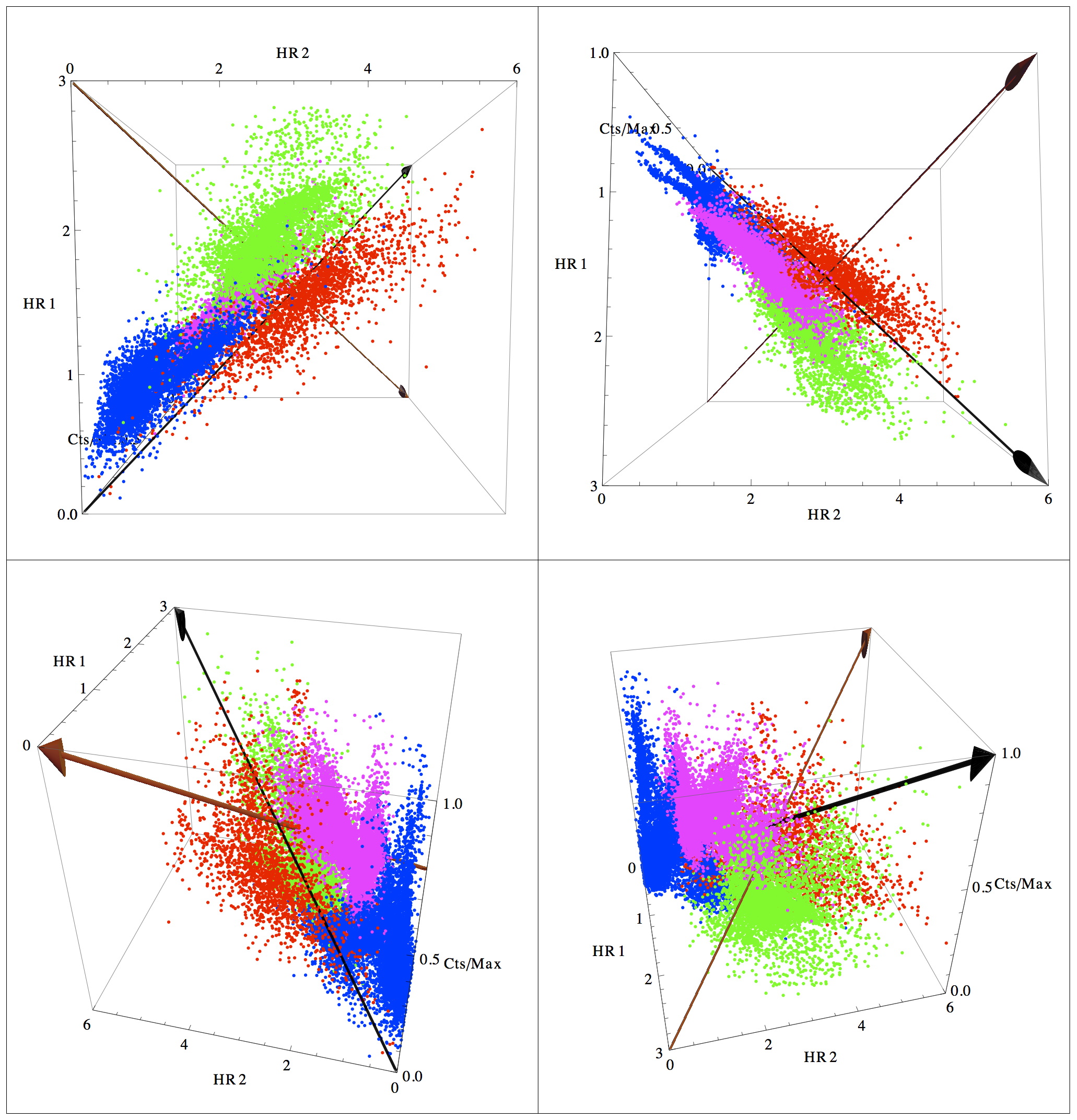}
\caption{
Classic BHs in blue, GRS1915-like BHs in green, pulsars in red,
non-pulsing NS in magenta.
Brown arrows shows direction of
increasing magnetic field strength and black arrow shows direction
of increasing mass accretion rate. 
The third axis
may be a measure of the Alfven radius for NS and
the ISCO for a BH as defined by Massi \& Bernando (2008).
}
\label{defaxis}
\end{figure*}

The parameters that may be
expected to influence X-ray emission (mass accretion
rate, binary separation, orbit inclination, mass ratio, magnetic
field strength, column density, etc.) are well established for many
systems.
Our CCI diagrams offer some clues as to how different
parameters may relate to location in phase space:
For example, Homan~\etal(2010) conclude
that the variety of behavior observed in atolls and Z-sources
 can be linked
to mass-accretion rate increasing from atolls to Z-sources.
MB08 have mass-accretion rate increasing from
atolls to Z-sources
(their Figure 3).
Assuming this to be correct we plot bursters, atolls, and Z-source
on one figure (Fig. \ref{zsatollbur})
and find that mass-accretion rates increases
on a diagonal from the corner with lowest
values of HR1, HR2, and Intensity to the corner with high HRs and
Intensity (Fig. \ref{defaxis}).
Note this implies that X-ray intensity within the ASM energy range
is {\it not} a direct indication
of mass accretion rate in agreement with our past work
(Vrtilek~\etal~1990, 1991, 1994).

Since the ASM is not very sensitive to NH we extracted the PCA 
data of sample sources, modeled the data, and adjusted the model for 
NH to see the effects.  Adjusting for the nominal NH (6.2e22cm$^{-2}$) 
to GRS 1915+105 
does place it closer to the other BH systems, but since the
other BHs also move slightly there is still a separation between GRS1915+105
and all other BHs.  But this
is a model dependent result.  The fact remains that the ASM raw data, 
uncorrected for NH, distinguishs four categories of sources: 
NS pulsing, NS non-pulsing, BH classical, and BH GRS1915-like. 
Detailed analysis of PCA data of individual sources using CCI is underway 
(Peris~\etal~2012; Buchan~\etal~2012; Cechura~\etal~2012).

\begin{figure*}
  \vspace*{174pt}
\includegraphics[width=6.2in,angle=0]{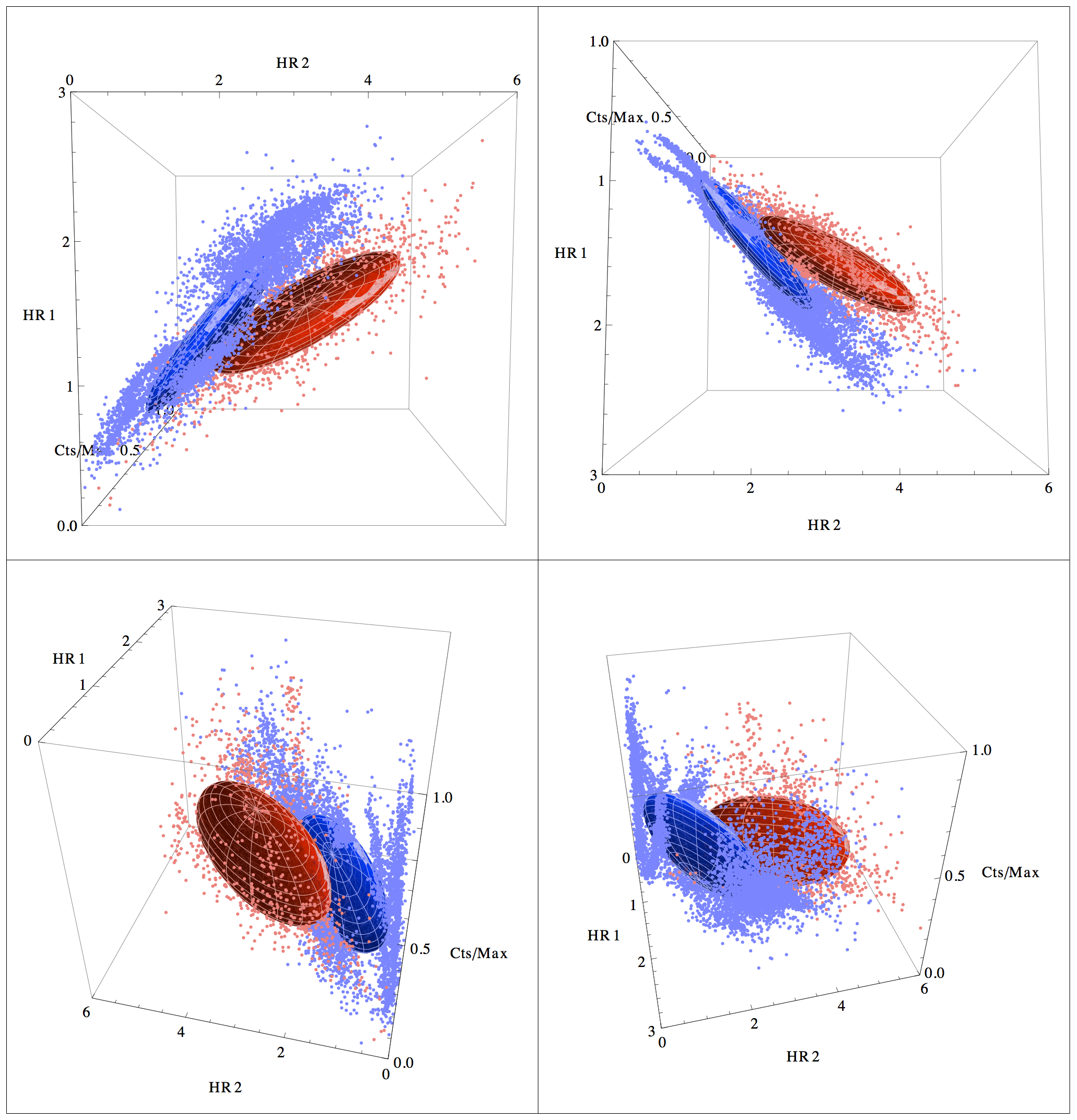}
\caption{
Four views of a CCI diagram with sources showing resolved jets
in blue and pulsars in red.
If we accept MB08 that pulsars cannot produce jets then we can
define the region were pulsars exist as a jet exclusion region.
50\% ellipsoids with all jet sources in blue and pulsars in red.
See Table 2 for values of the centroids.
}
\label{jetsourcespul}
\end{figure*}

MB08 also suggest that pulsars, distinguished by very
high magnetic fields, are clearly separated from
all other classes (their Fig. 1).
Migliari \& Fender (2006; hereafter MF06) suggest
a progression of source type associated with
differing magnetic field strengths.
Particularly interesting is that the
distribution of the sources
in our CCI diagram
appear to follow
the order delineated by MF06 for decreasing magnetic
field strength. This order suggests that
magnetic field strength in our CCI diagrams decreases diagonally
from the corner
with high HR2,
low HR1, and low I towards the corner with low HR2, high HR1,
and low I.
So in the upper left we have sources with the strongest magnetic fields,
the pulsars; going away from the pulsars towards the right we have first the
Z sources and then BH systems.

Mass accretion rate and magnetic field strength
represent two dimensions of our 3-D diagrams.
MB08 agree with Homan~\etal(2010) that mass accretion rate
increases from atoll to Z sources.
They agree with MF06 and Fender~\etal~(1997) that
X-ray pulsations and radio emission (taken as a measure
of jet strength) are strongly anti-correlated.
Following the work of MB08 we suggest that the third axis
can be represented by the ratio of the Alfven radius to
surface of the NS and
the innermost stable circular orbit (ISCO) for BHs.
For both BHs and NSs, we expect the inner
accretion disc to be an important contributor to the spectrum.
In the case of accreting NSs, we expect the inner
radius will be related to the Alfven radius, which depends on
the mass accretion rate and on the NS magnetic field.
For the BHs, the emitting radius may scale with the
general relativistic innermost stable circular orbit, which
depends on the BH mass, spin, and spin sense (prograde or
retrograde with the disc). The NS systems will have
an additional source of emission as gas disrupted by the magnetic
field near the Alfven radius is channeled onto the NS surface
(for stronger magnetic fields, the channeling will direct the
accretion onto a smaller area about the poles).
Our CCI diagrams may be related to MB08's physical
model by a transformation of coordinates.

\begin{table}
 \centering
 \begin{minipage}{140mm}
  \caption{Unclassified Objects }
\begin{tabular}{ll}
\hline
\hline
Object Name &Our classification\\
\hline
\hline
Circinus X-1&Non-pulsing neutron star\\
1700-37&Pulsar with massive companion\\
GX3+1&Non-pulsing neutron star\\
Cygnus X-3&Shares space with\\ 
&GRS1915-like and pulsars\\
\hline
\hline
\end{tabular}
\end{minipage}
\end{table}

\subsection{Defining the Jet Locus?}

MF06 suggest a
progression of source type and
activity correlated with the
detection of jets.  From most likely to least likely
to show jets, these sources range from BH systems with no intrinsic
magnetic fields,
low-mass NS systems with weak magnetic fields at high accretion (Z type),
low-mass NS systems with weak
magnetic fields at low accretion (atoll type),
and NS systems with strong magnetic fields (pulsars).
In general,
jet production is enhanced with mass accretion rate (as a means of
expelling angular momentum) but it is
inhibited by high magnetic fields on the surface of NSs or
in BH accretion discs.
MB08 quantified the progression
suggested by MF06 by considering
that in an accreting system
magnetic fields lines can be bent only when the magnetic pressure
is lower than the hydrodynamic pressure of the accreting material.
The basic condition for jet formation is then that the
location at which
the magnetic pressure and plasma pressure balance (Alfven radius)
is coincident with the NS surface or for a BH the innermost stable
orbit.

MB08 list 15 XRB systems that have resolved
jets and several additional systems with indirect evidence for
compact jets in the
form of a flat radio spectrum (seven of these are
bright enough in the ASM for this study).
The systems span a wide range of temporal and
spectral morphologies, and in roughly half the systems the nature
of the compact object is ambiguous.

Figure \ref{jetsourcespul} shows
systems containing both BHs and NSs
identified by MB08
as having resolved radio jets as well as the pulsars which according to MB08
cannot produce jets.  If we consider the progression
of source type and activity
correlated with the detection of jets suggested by Migliari \& Fender
(2006) we can imagine a curved plane
that separates pulsing systems from the others.  This plane
cuts through several sources
such as Cyg X-1 (blue), Sco X-1 (cyan), Cyg X-3 (magenta), and
GRS 1915+105 (pink).
We see that the progression is from
upper left towards lower right: in the upper left are the pulsars with
strong magnetic fields; going away from the pulsars toward the right
we have first the atoll and Z sources (differentiated mainly by
mass accretion rate) and then with the lowest magnetic fields the
BH systems.
This suggests that we can define a locus of our CCI space which
contains
systems that can form jets.
The ``jetline" as defined by
Fender, Belloni, and Gallo (2005) is for us a curved plane that
depends, not surprisingly, on
a combination of magnetic field strength, Alfven radius or ISCO, and
mass accretion rate. Since the ISCO depends on the spin parameter
of the BH, this interpretation is also consistent with the recent connection
made between BH spin period and jet power by Narayan \& McClintock (2011).
One of our ongoing projects is to define this
``jet plane" on the CCI diagram (Boroson~\etal~in preparation).

\section{Summary and Future Work}

We find that separate classes of XRBs fall into
distinct regions on a 3-dimensional colour-colour-intensity (CCI) diagram.
This provides a simple, model-independent method of distinguishing
between systems that contain BHs or NSs, and between systems
that contain pulsing and non-pulsing NSs.
CCI diagrams may also help us to define the observable characteristics
of systems that produce
jets and localize a ``jet plane"
that separates systems that cannot produce jets from those that can.

A hint of the physics underlying this
separation is provided by Massi \& Bernado (2008) who
characterize the systems in terms of their mass-accretion
rate, their magnetic field strengths, and the location
of the Alfven radius for NSs or the ISCO for BHs.
Given that the loci of different classes of objects
in CCI space is determined by intrinsic properties of the systems,
using
our prior knowledge of mass accretion rates and magnetic field
strengths we can define the directions of increasing accretion rate
and field strength.  Other factors (e.g., binary separation, mass ratio)
that may play roles are yet to be explored.

The ASM on RXTE has continuously monitored over 500 sources for the
past 15 years.  Only 200 of these sources are XRBs.  The rest include
a variety of bright X-ray emitters such as blazers, quasars, Seyfert
galaxies, cataclysmic variables, and Wolf-rayet stars.
It would be interesting to see if methods similar to those shown here
 can be used to
distinguish between different classes of cataclysmic variables (novae,
dwarf novae, polars, AM Can Ven) or galatic nuclei (blazers, quasars,
Seyferts of types 1 and 2).
Applying CCI techniques
to a variety of active galaxies may provide another
means of
probing the
microquasar/quasar connection.

Although the RXTE/ASM is no longer operating, MAXI (Monitor of ALL-sky
X-ray Image; Matsuoka~\etal~2009) operating on the Space Station 
measures the X-ray fluxes 
of over 1,000 X-ray sources (twice the number detected by the RXTE/ASM) over
the energy range 1-30keV
once every 96 minutes over the entire sky.
Such data should allow us to extend our CCI analysis to
many more sources, and enable us to study them out
to significantly higher
energies.
The Chandra X-ray observatory
has the demonstrated ability to detect XRBs in galaxies
other than our own.  M31 has been very well monitored by Chandra and
may present an opportunity to apply CCI techniques to XRBs in an external galaxy.
\\

\section*{Acknowledgments}

We would like to thank the RXTE/ASM team for easily and  
widely available ASM data.  We would like to thank Jeff 
McClintock and Josh Grindlay for pointing out possible
complications with the ASM data and Al Levine and Ron 
Remillard for informing us how to mitigate the problems. 
We thank Rosanne Di Stefano, John Raymond, and an anonymous 
referee for insightful comments and suggestions that greatly 
improved the quality and clarity of this paper.

\section{References and Citations}

\noindent
Belloni, T., Klein-Wolt, M., Mendez, M., van der Klis, M.,
\& van Paradijs, J. 2000, A\&A, 355, 271.

\noindent
Buchan, S., Peris, C., Boroson, B.S., \& Vrtilek, S.D. 2012,
(PCA observations of Cyg X-1), in preparation.

\noindent
Cechura, J., Peris, C., McCollough, M., \& Vrtilek, S.D. 2012, (PCA 
observations of Cyg X-3), in preparation.

\noindent
Fender, R.P., Belloni, T., \& Gallo, E. 2005, AP\&SS, 300,1.

\noindent
Fender, R.P.~\etal~2004, Nature, 427, 222.

\noindent
Hasinger, G., \& van der Klis, M. 1989, A\&A, 225, 79.

\noindent
Homan J.~\etal~2010, 719, 201.

\noindent
Koljonen, K.I.I., Hannikainen, D.C., McCollough, M.L., Pooley, G.G.,
\& Trushkin, S.A. 2010, MNRAS, 406, 307.

\noindent
Levine,A.M.~\etal~1996, ApJ, 469, 33.

\noindent
Liu, Q.Z., van Paradijs, J., \& van den Heuvel, E.P.J.
2001, A\&A, 368, 1021. (Lvv01)

\noindent
Liu, Q.Z., van Paradijs, J., \& van den Heuvel, E.P.J.
2000, A\&A Supplement Series, 147, 25.
(Lvv00)

\noindent
Matsuoka, M.~\etal~2009, PASJ, 61, 999.

\noindent
Massi, M, \& Bernado, M.K. 2008, A\&A,477,1.
(MB08)

\noindent
Migliari, S., \& Fender, R.P., 2006, MNRAS, 366, 79.

\noindent
Mirabel, I.F., \& Rodriguez, L.F. 1994, Nature, 371, 46.

\noindent
Narayan, R., \& McClintock, JEM.  2011 (presented at a
lunch talk at the CfA, paper is in press).

\noindent
Peris, C., Boroson, B.S., \& Vrtilek, S.D. 2012, (PCA
observations of GRS 1915+105), in preparation.

\noindent
Remillard, R., \& McClintock, J. 2006,
ARA\&A, 44, 49.
(RM06)

\noindent
Vrtilek, S.D.~\etal~1990, A\&A, 235, 162.

\noindent
Vrtilek, S.D.~\etal~1991, ApJ, 376, 278.

\noindent
Vrtilek, S.D.~\etal~1994, ApJL, 436,9.

\noindent
Wijnands, R.~\etal~1998, ApJ, 504, L35.

\label{lastpage}

\end{document}